\documentclass[fleqn,usenatbib]{mnras}

\usepackage{newtxtext,newtxmath}
\usepackage{scrextend}

\usepackage[T1]{fontenc}

\DeclareRobustCommand{\VAN}[3]{#2}
\let\VANthebibliography\thebibliography
\def\thebibliography{\DeclareRobustCommand{\VAN}[3]{##3}\VANthebibliography}

\usepackage{graphicx}	
\usepackage{amsmath}	

\title[Resolved outflow metallicity measurements]{DUVET: Resolved direct metallicity measurements in the outflow of starburst galaxy NGC~1569}

\author[M. Hamel-Bravo et al.]{
Magdalena J. Hamel-Bravo,$^{1, 2}$
Deanne B. Fisher,$^{1, 2}$
Danielle Berg,$^{3}$
Bjarki Björgvinsson,$^{9,10}$\newauthor
Alberto D. Bolatto,$^{4}$
Alex J. Cameron,$^{5}$
John Chisholm,$^{3}$
Drummond B. Fielding,$^{6}$
Rodrigo Herrera-Camus,$^{7}$\newauthor
Glenn G. Kacprzak,$^{1,2}$
Miao Li,$^{8}$
Barbara Mazzilli Ciraulo,$^{1,2}$
Anna F. McLeod,$^{9,10}$
Daniel K. McPherson,$^{1, 2}$\newauthor
Nikole M. Nielsen,$^{1,2}$
Bronwyn Reichardt Chu,$^{1,2,9,10}$
Ryan J. Rickards Vaught,$^{11}$
and Karin Sandstrom$^{11}$
\\
$^{1}$Centre for Astrophysics and Supercomputing, Swinburne University of Technology, Hawthorn, Victoria 3122, Australia\\
$^{2}$ARC Centre of Excellence for All Sky Astrophysics in 3 Dimensions (ASTRO 3D), Australia\\
$^{3}$Department of Astronomy, University of Texas, Austin, TX 78712, USA\\
$^{4}$Department of Astronomy, University of Maryland, College Park, MD 20742, USA\\
$^{5}$Sub-department of Astrophysics, University of Oxford, Keble Road, Oxford OX1 3RH, United Kingdom\\
$^{6}$Center for Computational Astrophysics, Flatiron Institute, 162 Fifth Avenue, New York, NY 10010, USA\\
$^{7}$Departamento de Astronom\'ia, Universidad de Concepci\'on, Barrio Universitario, Concepci\'on, Chile\\
$^{8}$Department of Physics, Zhejiang University, 866 Yuhangtang Road, Hangzhou, 310058, China\\
$^{9}$Centre for Extragalactic Astronomy, Department of Physics, University of Durham, South Road, Durham DH1 3LE, UK\\
$^{10}$Institute for Computational Cosmology, Department of Physics, University of Durham, South Road, Durham DH1 3LE, UK\\
$^{11}$Department of Astronomy \& Astrophysics, University of California, San Diego, 9500 Gilman Drive, MC0424, La Jolla, CA, 92093 USA\\
}

\date{Accepted XXX. Received YYY; in original form ZZZ}

\pubyear{2024}

\begin{document}
\label{firstpage}
\pagerange{\pageref{firstpage}--\pageref{lastpage}}
\maketitle

\begin{abstract}
We present the results of direct-method metallicity measurements in the disk and outflow of the low-metallicity starburst galaxy NGC~1569.  We use Keck Cosmic Web Imager observations to map the galaxy across 54$\arcsec$ (800~pc) along the major axis and 48$\arcsec$ (700~pc) along the minor axis with a spatial resolution of 1$\arcsec$ ($\sim$15~pc). We detect common strong emission lines ([\ion{O}{III}]~$\lambda$5007, H$\beta$, [\ion{O}{II}]~$\lambda$3727) and the fainter [\ion{O}{III}]~$\lambda$4363 auroral line, which allows us to measure electron temperature ($T_e$) and metallicity. Theory suggests that outflows drive metals out of the disk driving observed trends between stellar mass and gas-phase metallicity. Our main result is that the metallicity in the outflow is similar to that of the disk, $Z_{\rm out} / Z_{\rm ISM} \approx 1$. This is consistent with previous absorption line studies in higher mass galaxies. Assumption of a mass-loading factor of $\dot{M}_{\rm out}/{\rm SFR}\sim3$ makes the metal-loading of NGC~1569 consistent with expectations derived from the mass-metallicity relationship. 
Our high spatial resolution metallicity maps reveal a region around a supermassive star cluster (SSC-B) with distinctly higher metallicity and higher electron density, compared to the disk. Given the known properties of SSC-B the higher metallicity and density of this region are likely the result of star formation-driven feedback acting on the local scale. 
Overall, our results are consistent with the picture in which metal-enriched winds pollute the circumgalactic medium surrounding galaxies, and thus connect the small-scale feedback processes to large-scale properties of galaxy halos.

\end{abstract}




\section{Introduction}

Galaxy scale winds (or outflows) play a key role in galaxy evolution by redistributing baryons from the disk of star-forming galaxies \citep[e.g.][]{Chevalier1985, Heckman1990}. Galactic winds are widely observed in star-forming galaxies \citep{Veilleux2005,Veilleux2020}, and are frequently invoked in galaxy evolution theory to explain a number of observations, such as the stellar mass function of galaxies \citep[e.g.][]{Pillepich2018} and the metal content of the circumgalactic medium \citep[e.g.][]{Tumlinson2017,Peroux2020}. Despite their critical importance for theory, properties of outflows, like the metallicity and electron density, remain unconstrained. 


Outflows likely play a key role in redistributing metals in galaxies. All metals are created during the many stages of stellar evolution, in the galaxy disk, and yet metals make up a significant mass component of the gaseous halos of galaxies \citep[e.g.][]{Werk2013}. Outflows containing metals originating from supernovae ejecta and entrained ISM gas offers an obvious mechanism to enrich halos \citep{Peeples2011, Christensen2018}. Moreover, the relationship between total galaxy stellar mass and metallicity (MZR) shows that lower-mass galaxies have lower metallicities \citep[e.g.,][]{Tremonti2004, Berg2012, Kirby2013}. Most galaxy evolution models and simulations incorporate stellar feedback \citep[e.g.][]{Springel2003, Oppenheimer2006, Hopkins2014} as a mechanism to decrease ISM metallicities in low-mass galaxies. This process is quantified by the metal-loading factor, $\zeta$, which describes how efficient outflows are at removing metals from the ISM relative to how many metals are retained due to star-formation \citep{Peeples2011}: 
\begin{equation}
    \label{eq:metalloading}
    \zeta = \frac{Z_{\rm out}}{Z_{\rm ISM}} \frac{\dot{M}_{\rm out}}{\rm SFR}.
\end{equation}

\noindent $Z_{\rm out}$ and $Z_{\rm ISM}$ are the outflow and ISM metallicities, respectively, $\dot{M}_{\rm out}$ is the mass outflow rate, and SFR is the star formation rate. Observational constraints on $Z_{\rm out}/Z_{\rm ISM}$ of galaxies across a wide range of galaxy properties are needed to determine if this process is sufficient to shape the MZR.  


To measure $Z_{\rm out}/Z_{\rm ISM}$, one must measure the metallicity in both the ISM and the outflow from the galaxy. While substantial literature exists characterizing the gas-phase metallicity within galaxies and the mass-loading of outflows \citep[reviewed in][]{Veilleux2020}, there is a lack of constraints on the metal content outside the galactic midplane. Gas-phase metallicity measurements are usually determined using optical emission lines to estimate the O/H abundance of the gas as a proxy for metallicity. 

The ``strong line'' method \citep[e.g.,][]{Maiolino2019, Kewley2019} estimates metallicity indirectly using the ratio of strong emission lines (e.g.,  [(\ion{O}{II}]~$\lambda$3727~+~[\ion{O}{III}]~$\lambda$5007)~/~H$\beta$), calibrated either empirically to \ion{H}{II} regions in disks or using photoionization models. We must note that metallicities measured from different methods have been shown to differ up to 0.7 dex \citep{Kewley2008, Maiolino2019}. The strong-line method is very useful for estimating metallicities in the ISM; however, it is unsuitable for measuring metallicity in outflows because it is calibrated primarily in \ion{H}{II} regions where physical conditions differ substantially from outflows. Indeed, using strong-lines may not only be unreliable, but would lead to misleading outflow metallicities.

The ``direct method'' uses the ratio of auroral to strong lines (like [\ion{O}{III}]~$\lambda$4363/ [\ion{O}{III}]~$\lambda$5007) to measure the electron temperature ($T_e$) of the gas. It is then possible to infer metallicity, assuming that metals dominate the cooling of the gas \citep{Berg2012, Perez-Montero2017}. In this method, minimal assumptions on the photoionization source are made. $T_e$-based metallicities are, therefore, a more reliable way to determine outflow metallicities. Auroral lines, however, are typically 50--100 times fainter than H$\beta$, making them challenging to observe in intrinsically faint outflows. Moreover, they are only sufficiently bright in high-temperature / low-metallicity environments. Therefore, while the direct method does open a window to measuring the metallicity of outflows, it is only feasible in low-metallicity galaxies. 

Alternatively, outflow metallicity can be inferred via X-rays of the hot phase of the outflowing gas. The hot phase gas has been observed for both NGC~1569 \citep{Martin2002} and M~82 \citep{Lopez2020} to be significantly metal-enriched by a factor of $\sim3.5\times$ that of the disk. This is expected as hotter gas in outflows likely originates from the supernovae ejecta, while colder phases more likely probe entrained gas from the ISM.  
The colder phases of the outflow can be probed via metal absorption lines. \cite{Chisholm2018} use this method to determine the metallicity in the outflow of seven galaxies with stellar masses spanning $10^7-10^{11} M_\odot$. Using direct-method gas-phase abundances, they found an inverse correlation between $Z_{\rm out}/Z_{\rm ISM}$ and stellar mass, showing that outflows in low-mass galaxies are more enriched in relation to their host galaxy. They show that these results can reproduce the MZR, under a set of assumptions described in that work.

More recently, \cite{Cameron2021} pioneered a self-consistent study of the metal content throughout the baryon cycle using direct-method abundances in the disk, inflow, and outflow of Mrk~1486. They observed an $Z_{\rm out}/Z_{\rm ISM}\approx1.6$.
This emission line metal ratio is only half that of the absorption-line value from \cite{Chisholm2018} for the same galaxy. 
This would suggest that absorption-line methods may overestimate $Z_{\rm out}$ relative to emission-line based metallicity, however, the systematics of comparing absorption-line to emission-line methods remain poorly known. Clearly, more work is needed to robustly constrain the metallicity of outflows. 

In this work, we extend the pioneering method of \citet{Cameron2021} to galaxy NGC~1569. NGC~1569 is a well-studied, nearby (3.25 Mpc, \citealp{Tully2013}), low metallicity (12+log(O/H) = 8.19, \citealp{Kobulnicky1997}), low mass (log($M_{\star}/M_{\odot}$ = 8.6, \citealp{Leroy2019}), starburst galaxy. It has two supermassive star clusters (SSC) that formed after a period of high star formation approximately 10 Myrs ago \citep{Hunter2000}. H$\alpha$ emission shows a complex structure around the SSCs, with filaments and bubbles suggesting a stellar feedback driven outflow \citep{Heckman1995, Westmoquette2007}. Analysis of the \ion{H}{I} kinematics suggests a mass loading of the cold gas in the wind of $\dot{M}_{\rm out}/{\rm SFR}\sim 3$ \citep{Johnson2012,McQuinn2019}. 
X-ray observations show evidence of a metal enriched outflow \citep{Martin2002}. Due to its proximity, low gas-phase metallicity, and observed metal-enriched outflows, NGC~1569 provides an excellent laboratory for resolving the metallicity of a galactic wind. 
 
The remainder of this paper is organized as follows:
We present Keck/KCWI observations of NGC~1569 in Section~\ref{sec:obs}. Section~\ref{sec:analysis} describes our spectral analysis,  including our line-fitting method (\S\ref{sec:linefits}), emission-line maps (\S\ref{sec:linemaps}), and separation of the disk and outflow regions (\S\ref{sec:regions}). We then use the emission-line maps to examine the electron temperature and metallicity maps in Section~\ref{sec:metallicity} and discuss possible temperature fluctuations in Section~\ref{sec:fluctuations}. We use these results to examine the outflow of NGC~1569, focusing on the $T_e$ fluctuations in Section~\ref{sec:enrichment}, the density in Section~\ref{sec:density}, and metal loading in Section~\ref{sec:loading}. Finally, we present our summary in Section~\ref{sec:summary}. We assume a flat $\Lambda$CDM cosmology with
$H_{0}=69.3$~km~Mpc$^{-1}$~s$^{-1}$, $\Omega_{m}=0.3$, and $\Omega_{\Lambda}=0.7$.

\section{Observations and data reduction}\label{sec:obs}

\begin{figure*}
    \centering
    \includegraphics[width=0.8\textwidth]{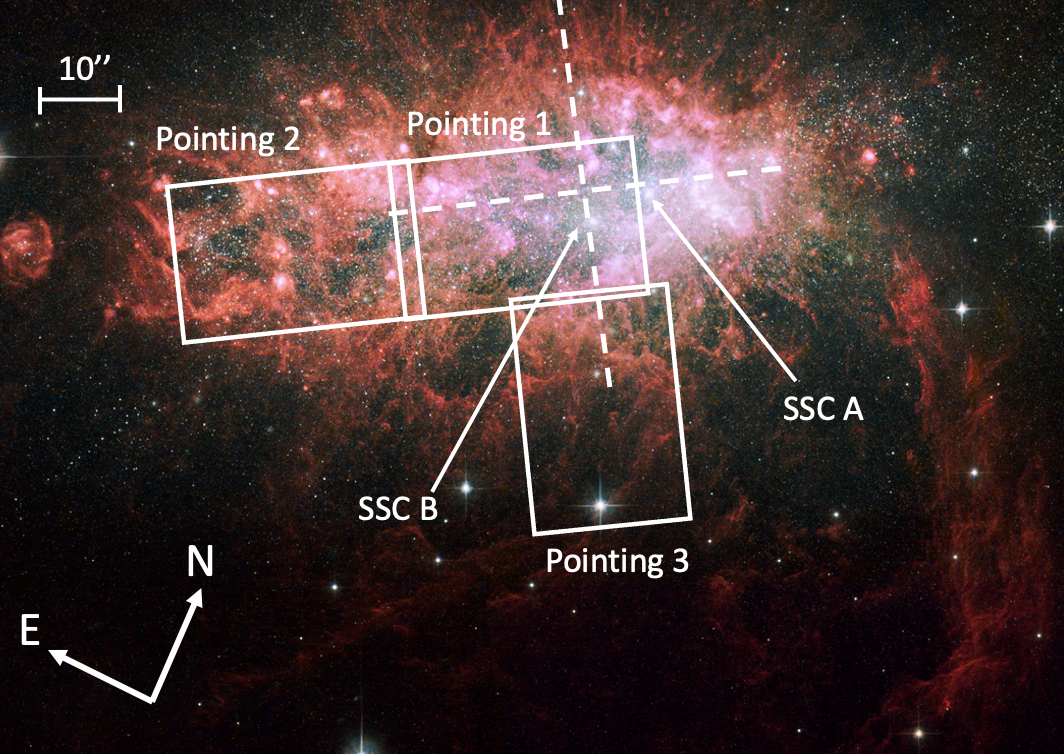}
    \caption{HST composite image of NGC~1569, red: F658N filter (H$\alpha$ + [\ion{N}{II}]), green: F606W filter, light blue: F5002N filter ([\ion{O}{III}]) and dark blue: F487N filter (H$\beta$). Plotted on top of the HST image are our three KCWI positions. Our KCWI data maps 54\arcsec ($\sim$800~pc) across the major axis and 48\arcsec ($\sim$700~pc)($\sim$ across the minor axis. The arrows show the position of the two known super star clusters (SSC) SSC~A and SSC~B. The two dashed lines show the major and minor axis of the galaxy obtained from {\it Spitzer}/IRAC 3.6$\mu$m band observations.}
    \label{fig:pointings}
\end{figure*}

We observed NGC~1569 at three different positions with Keck/KCWI, as part of the Deep near-UV observations of Entrained gas in Turbulent galaxies (DUVET) survey. We targeted two positions aimed at the galaxy disk, and a third position towards the minor axis. The minor axis frame is co-located with an H$\alpha$ filament. Figure~\ref{fig:pointings} shows an HST composite image.  The white dashed lines in the figure illustrate the position of the major and minor axes, which intersects at the center of {\it Spitzer}/IRAC 3.6$\mu$m band observations. We also show the position of our three KCWI pointings. NGC~1569 has an inclination of 60\textdegree\, measured from \ion{H}{I} observations \citep{Stil2002}. Typical of dwarf galaxies, the light of the galaxy is not symmetric. The derived inclination may, therefore, have significant uncertainty. Our field-of-view covers a projected distance of $\sim$700~pc along the minor axis. Considering an inclination of 60\textdegree\ this corresponds to a distance of $\sim$810~pc from the galaxy midplane. If we consider that the galaxy is less inclined, 50\textdegree\ for example, the distance could be $\sim$915~pc. Through out this work, we report projected distances. 

The KCWI observations were conducted on November 14th, 2020, with a mean seeing of 1.2\arcsec. All observations used the large IFU slicer. The field of view of each individual frame is 20\arcsec~$\times$~33\arcsec with a spaxel size of 0\farcs29~$\times$~1\farcs35. We used two different configurations of the blue medium-dispersion grating to observe each pointing. The first setting is centered at 4050~{\AA} (``blue'' setting) and the other at 4700~{\AA} (``red'' setting). These configurations gave us a total continuous wavelength range of 3600~{\AA} -- 5130~{\AA}, with a spectral resolution of R$\sim$2000. We used half-slice dither in both configurations to increase the spatial sampling. Our observation strategy and reduction methods are described in the literature for other DUVET targets \citep{Cameron2021,McPherson2023}. 

To avoid saturation on bright emission lines such as [\ion{O}{III}]~$\lambda$5007, each position was observed with short exposures while long exposures were used to detect fainter lines like [\ion{O}{III}]~$\lambda$4363. Pointing~1 was observed with seven exposures in the red setting (3 $\times$ 30~s,  3 $\times$ 100~s, and 1 $\times$ 300~s) and four in the blue setting (1 $\times$ 30~s and 3 $\times$ 100~s). Pointing~2 was observed with six exposures in the red setting (3 $\times$ 30~s and 3 $\times$ 300~s) and four in the blue setting (1 $\times$ 100~s and 3 $\times$ 300~s). Pointing~3 was observed with six exposures in the red setting (3 $\times$ 30~s and 3 $\times$ 300~s) and three in the blue setting (3 $\times$ 300~s). We also obtained sky fields for the blue and red settings of 300~s exposure each. 

To reduce observations we used the KCWI Data Extraction and Reduction Pipeline v1.1.0\footnote{\url{https://github.com/Keck-DataReductionPipelines/KcwiDRP}}. To ensure clean spectra from the galaxy, we executed a sky subtraction using the sky frame previously described. Offset skies are scaled to match science exposure times. Flux calibration was performed within the pipeline using the standard star Feige25. Prior to combining individual exposures, we aligned individual cubes for each pointing by minimizing the residual in the H$\gamma$ emission line, which is observed in both the red and blue settings and in all pointings. This was carried out to account for any possible small shifts in the spatial coordinates. We combined long and short exposures using the [\ion{O}{III}]~$\lambda$5007/$\lambda$4959 ratio as an indicator of saturation. Assuming a value of 3 \citep{Osterbrock2006} we determine spaxels for which [\ion{O}{III}]~$\lambda$5007 was saturated. For those spaxels, we replaced [\ion{O}{III}]~$\lambda$5007, [\ion{O}{III}]~$\lambda$4959 and H$\beta$ in long exposures with short exposures in a 20~{\AA} wavelength region centered on the emission line. This procedure is discussed in more detail in \cite{McPherson2023}. 

Exposures were reprojected into 0.29\arcsec\ $\times$ 0.29\arcsec\ spaxels and co-added using the python package \textsc{Montage}\footnote{\url{http://montage.ipac.caltech.edu/}}. We then binned our cubes in 3 $\times$ 3 spaxels to have a spaxel size of 0.87\arcsec $\times$ 0.87\arcsec, which is approximately half of the FWHM of our point spread function. After co-addition and re-binning of frames, several small differences between red and blue settings and positions needed to be accounted for to generate a uniform data set. 

First, a small difference in continuum surface brightness was present between the red and blue settings. This is likely due to imperfections in flux calibration. We scaled the blue spectrum in each spaxel by the average ratio observed in the overlapping wavelength range between the two gratings, which was between 4250~{\AA} and 4500~{\AA}. This scaling was of order $\sim$1\% of the flux. We then combined blue and red cubes for each pointing. 

Second, to spatially align the three pointings shown in Fig.~\ref{fig:pointings}, we used H$\gamma$ flux in the overlapping regions between the pointings. We iteratively shift the position of each pointing by integer spaxels in all directions and minimize the residual of $\Delta f_{\rm H\gamma}$ between the two positions. 

Finally we account for small differences in fluxes between pointings. Pointing~1 was observed directly after the flux standard. We, therefore, assume it is best calibrated pointing. We compared the shape of the continuum in the overlapping regions between Pointings~1 and 2, as well as between Pointings~1 and 3. We then fitted a polynomial to the average shift of all overlapping spaxels and shifted the spectra in all spaxels in Pointing~2 and Pointing~3 by the corresponding polynomial. This change is likewise of order $\sim$1\%, or less, of the continuum flux.

\subsection{Continuum subtraction}

Milky Way extinction is corrected using the \cite{Cardelli1989} extinction law, with extinction $A_v = 1.85$  \citep{Schlafly2011}. Then we corrected for the stellar continuum. For continuum fitting we implement the standard method via pPXF \citep{Cappellari2017}, with model spectra sourced from the Binary Population and Spectral Synthesis code v2.3 (BPASS v2.3). The BPASS templates contain single and binary stars with an initial mass function extending between 0.1 and 300~M$_{\odot}$. Prior to running the continuum fit, we masked all visually identifiable emission lines in a 200~km~s$^{-1}$ window. 

Our observations include both regions with significant stellar continuum (Pointing~1 \& 2) and those with minimal contribution from the continuum (Pointing~3). Modelling the stellar continuum is a non-linear process, and to do so in regions of low S/N would likely result in incorrect adjustments to the spectrum. To determine which spaxels have sufficient stellar continuum, we computed the continuum signal-to-noise ratio (S/N) in a band of 200~\AA, ranging from 4600~\AA\ to 4800~\AA. While faint emission exists, the integrated flux in this band is heavily dominated by continuum. The fitting was performed only in spaxels where the signal-to-noise ratio (S/N) of the continuum was greater than 3. For spaxels with a continuum signal-to-noise ratio lower than 3, we fit a constant value to represent the continuum.

\section{Spectral analysis}\label{sec:analysis}

\subsection{Emission line fitting}\label{sec:linefits}

For the emission line fitting we used our internal software, \texttt{threadcount}\footnote{\url{https://github.com/astrodee/threadcount}}, built on work done by \cite{ReichardtChu2022a}. This software allows for the user to input multiple models with different numbers of Gaussians and different constraints. The software automatically selects the best model based on the Bayesian Information Criterion (BIC), given a $\Delta$BIC threshold. To calculate the uncertainties in emission line fluxes, \texttt{threadcount} employs Monte Carlo simulations. It generates simulated spectra by adjusting the observed flux in each spectral pixel using a normal distribution with a standard deviation derived from the observed variance in our data. The uncertainty in the flux is then computed as the standard deviation of the measured fluxes in the simulated spectra. For this study, 100 simulations for each spaxel were performed. 

We first fitted the H$\beta$ and H$\gamma$ emission lines using a single Gaussian model. We corrected for the galaxy internal dust extinction based on the H$\beta$/H$\gamma$ flux ratio. We assumed an intrinsic value of ${f_{\rm H\beta} / f_{\rm H\gamma} = 2.12}$, which is typical for gas at a temperature of about $10^4$~K \citep{Osterbrock2006}. As NGC~1569 is a low metallicity galaxy, we used the extinction law from the Large Magellanic Cloud from \cite{Gordon2003} with $R_v \approx$ 2.7. We found an average extinction of $A_v$ = 1.2 on the disk and an average of $A_v$ = 1.5 in the outflow. 

We fitted the H$\beta$ and [\ion{O}{III}]~$\lambda$5007 lines using single Gaussian models, we do not see evidence of double Gaussian emission. For the [\ion{O}{II}]~$\lambda\lambda$3726,9 doublet, we used a double Gaussian model with a constrained shift between the centroids but no constraint on the flux ratios.

The [\ion{O}{III}]~$\lambda$4363 line can be blended with emission from [\ion{Fe}{II}]~$\lambda$4360. This has been observed in stacked galaxy spectra and HII regions \citep{Curti2017, Berg2020, RickardsVaught2023}. We observe [\ion{Fe}{II}] in some spaxels in our KCWI data, but most spaxels do not show any significant detection.  The bottom panels of Fig.~\ref{fig:emissionmaps} show three examples of [\ion{O}{III}]~$\lambda$4363 emission lines from our data. In panel (a) we show an example where no [\ion{Fe}{II}] contamination is observed, panel (b) shows an example with a small [\ion{Fe}{II}] detection and panel (c) shows the strongest [\ion{Fe}{II}]~$\lambda$4360 detection.

In order to account for possible contamination by [\ion{Fe}{II}]~$\lambda$4360, we fitted the [\ion{O}{III}]~$\lambda$4363 emission with two models: a single Gaussian and a double Gaussian with a fixed shift between the centroids. We determined the best model by performing a BIC test. A $\Delta$BIC value of 10 is typically considered significant evidence against the single Gaussian model \citep{Kass1995, Swinbank2019}. However, for some astrophysical applications, a higher $\Delta$BIC may be more appropriate \citep{ReichardtChu2022a, ReichardtChu2024}. We tested two $\Delta$BIC thresholds: 20, 500. Using a threshold of 20, we detected [\ion{Fe}{II}] emission in $\sim$20\% of the spaxels with [\ion{O}{III}]~$\lambda$4363 detections, while using a threshold of 500, we detect [\ion{Fe}{II}] in $\sim$13\% of spaxels. An example of this is shown in Panel (b) of Fig.~\ref{fig:emissionmaps}. For this spaxel, we detect [\ion{Fe}{II}] when using a $\Delta$BIC of 20, but not when using a $\Delta$BIC of 500. From this point on, we will consider a $\Delta$BIC of 20. However, it is worth noting that increasing the $\Delta$BIC does not significantly impact the results. In the subsequent sections, we will discuss how the existence of [\ion{Fe}{II}]~$\lambda$4360 in the vicinity of the [\ion{O}{III}]~$\lambda$4363 emission line introduces an element of uncertainty in our measurements.



\subsection{Emission Line Maps}\label{sec:linemaps}

\begin{figure*}
    \centering
    \includegraphics[width=1\textwidth]{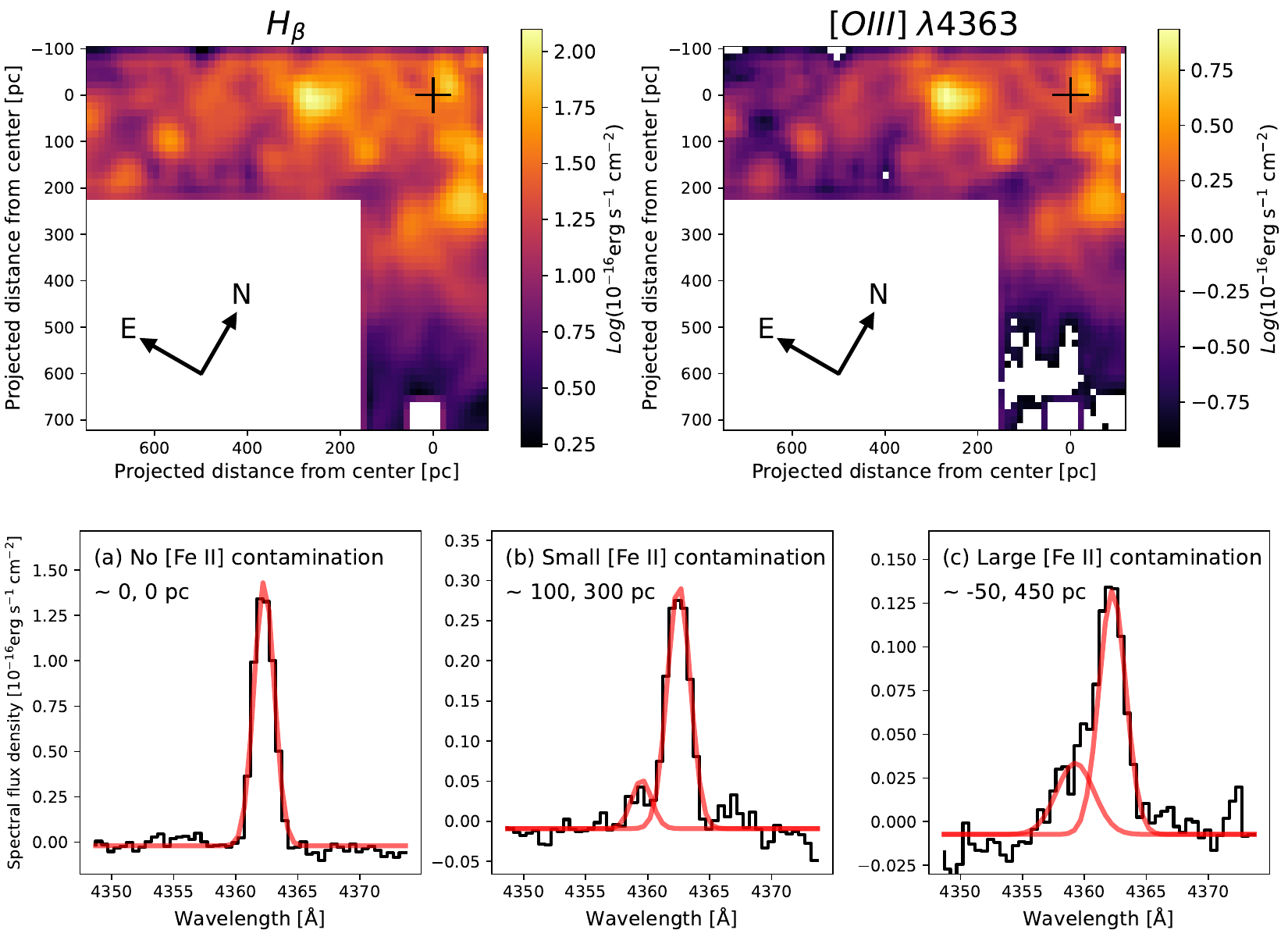}
    \caption{Top row: On the left, emission line flux map of H$\beta$ from KCWI. The x and y-axes show the position to the galaxy center, determined from {\it Spitzer}/IRAC 3.6$\mu$m band observations. We mark the center with a black $+$ symbol. We masked out a 4$\arcsec$ $\times$ 4$\arcsec$ region around a foreground star at (0, 700~pc). On the right, the emission line flux map of the [\ion{O}{III}]~$\lambda$4363 line. The spaxel size is 0.87\arcsec $\times$ 0.87\arcsec (13 $\times$ 13~pc). SSC~B is located near the emission line cavity at (0,200~pc) in the images. Bottom row: three examples of the [\ion{O}{III} ]~$\lambda$4363 emission line. Panel (a) shows a spaxel where we observe no [\ion{Fe}{II}]~$\lambda$4360 contamination, panel (b) shows a spaxel with small [\ion{Fe}{II}] contamination and panel (c) shows a spaxel with large [\ion{Fe}{II}] contamination. The position in the map for each example is indicated in the top right of each panel. In black we show the data and in red the fit using a $\Delta$BIC of 20. }
    \label{fig:emissionmaps}
\end{figure*}

Our KCWI observations show strong H$\beta$ emission at $>5\sigma$ level across the entire field. The top left panel of Fig.~\ref{fig:emissionmaps} shows a map of the H$\beta$ line flux, with values ranging from  ${\sim2-100 \times 10^{-16}}$~erg~s$^{-1}$~cm$^{-2}$. We have masked out a region around (0,700~pc) where a foreground star is present. We define the galaxy center by the peak luminosity in the {\it Spitzer}/IRAC 3.6$\mu$m band (P.I. Fazio, Program I.D. 69), as shown by the $+$ symbol in Fig.~\ref{fig:emissionmaps}.  {\it HST}/ACS H$\alpha$ images also exist of NGC~1569 (P.I. Aloisi, Program I.D. 10885), revealing complex, clumpy structures of ionized emission (see Fig.~\ref{fig:pointings}). We observe similar structures in H$\beta$ flux. In particular, we note a peak in H$\beta$ flux roughly 250~pc Southeast along the major axis of the disk coincident with an \ion{H}{II} region in the H$\alpha$ data; a cavity roughly 200~pc South of the major axis, where SSC~B is located; and a bright filament extending to 700~pc Southwest along the minor axis of the galaxy. This bright filament is particularly interesting to study outflows because of its association with outflowing gas caused by supernovae. The bright filament shows an H$\beta$ flux that is one order of magnitude higher than its surroundings.

We detect [\ion{O}{III}]~$\lambda$4363 at $>3\sigma$ level in 93\% of our spaxels, with an upper-limit in the flux of $\sim$0.1 $\times 10^{-16}$~erg~s$^{-1}$~cm$^{-2}$. The top right panel of Fig.~\ref{fig:emissionmaps} shows the  [\ion{O}{III}]~$\lambda$4363 emission line flux. The [\ion{O}{III}] flux has a peak of $\sim10^{-16}$~erg~s$^{-1}$~cm$^{-2}$ and decays by 2~orders-of-magnitude towards the minor axis to the edge of the KCWI pointing, $\sim700$~pc away from the galaxy center. The bright H$\beta$ filament can also be observed in [\ion{O}{III}]~$\lambda$4363 emission. As stated in section~\ref{sec:linefits} we observe contamination of [\ion{Fe}{II}]~$\lambda$4360 in $\sim$20\% of spaxels with [\ion{O}{III}]~$\lambda$4363 using a $\Delta$BIC of 20. The bottom panel in Fig.~\ref{fig:emissionmaps} shows three examples of [\ion{O}{III}]~$\lambda$4363 emission lines with different degrees of [\ion{Fe}{II}]~$\lambda$4360 contamination. We have determined the maximum [\ion{Fe}{II}]~$\lambda$4360 to [\ion{O}{III}]~$\lambda$4363 ratio to be $\sim$0.6 and the minimum $\sim$0.1. On average, the [\ion{Fe}{II}] flux is 0.2 times the [\ion{O}{III}]~$\lambda$4363 flux.

We find very high S/N detections of all strong-lines (e.g., [\ion{O}{ii}]~$\lambda\lambda$3727,9 and [\ion{O}{iii}]~$\lambda$5007) across all fields, all showing similar structure to H$\beta$ and [\ion{O}{III}]~$\lambda$4363 (see Appendix~\ref{sec:apen_emission_maps}). We can, therefore, make resolved measurements of $T_e$ and metallicity in both the disk and the filament that is leaving the galaxy to a distance of 700~pc from the midplane.



\subsection{Separating Disk And Outflow Regions}\label{sec:regions}
\label{sec:Disk_outflow}

\begin{figure}
    \centering
    \includegraphics[width=\linewidth]{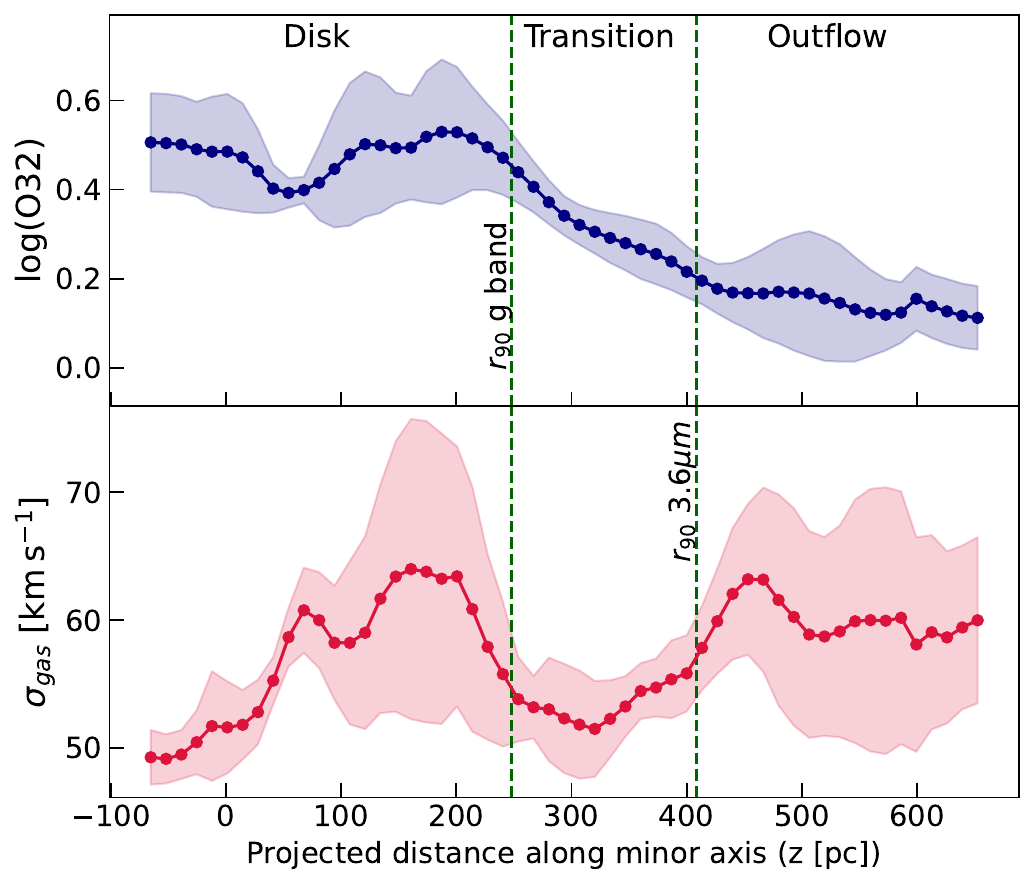}
    \caption{Ionization and velocity dispersion profiles extending above and below the disk midplane. Top panel in blue shows log(O32) vs. z, where z is the distance from the midplane of the galaxy in parsecs. Blue points show the mean value in vertical slices parallel to the midplane of the galaxy, and the shaded blue area shows the $\pm$1$\sigma$ around the mean. The bottom panel in red shows the velocity dispersion from the [\ion{O}{iii}]~$\lambda$5007 emission line. Red points show the mean value and the shaded red area shows the $\pm$1$\sigma$ around the mean. The green dashed vertical lines at z $\approx$ 250 pc and z $\approx$ 400 pc, show the distance from the midplane at which the integrated flux of PanSTARRS g-band luminosity profile and Spitzer/IRAC 3.6$\mu$m band luminosity profile reach 90\% respectively. At the top we label the three different regions as defined in section~\ref{sec:Disk_outflow}.}
    \label{fig:O32_profile}
\end{figure}

To determine the metal loading factor, we need to measure the metallicity in the disk and the outflow separately. The galaxy disk has no clear boundary, so it is not straightforward to define where our KCWI field of view is sampling the disk and where it is sampling the outflow.  NGC~1569 has been associated with outflows in several studies \citep[e.g.][]{Westmoquette2007, Johnson2012}, particularly because of the extended H${\alpha}$ filaments. Our KCWI observations were planned to sample both the disk and the outflow by placing two pointings parallel to the major axis of the galaxy, on top of the galaxy's midplane (pointing~1 and pointing~2 in Fig.~\ref{fig:pointings}) and one on top of a bright H${\alpha}$ filament associated with an outflow  along the minor axis (pointing~3 in Fig.~\ref{fig:pointings}).

With archival {\it Spitzer}/IRAC observations in the 3.6$\mu$m band and Pan-STARRS g-band \citep{Flewelling2020} observations we determine the vertical extent of the galaxy to define a region where our KCWI observations are sampling the ISM and the outflow. The 3.6$\mu$m-band is typically interpreted as a good proxy for total stellar mass in the disk, while the mass-to-light ratio of the g-band will be more affected by younger (<100~Myr) stars.  We measure the surface brightness profile along the minor axis of the galaxy and consider the brightest point to be the midplane of the disk. We determine the distance from the midplane at which the integrated surface brightness reaches the 50\% ($r_{50}$) and 90\% ($r_{90}$) of the total integrated surface brightness. The results for the IRAC 3.6$\mu$m band are: $r_{50,3.6\mu{\rm m}}\approx180$~pc and $r_{90,3.6\mu{\rm m}}\approx408$~pc. The results for the Pan-STARRS g-band are: $r_{50,g}\approx86$~pc and $r_{90,g}\approx247$~pc. This shows that the young stars are distributed more compactly than the older stars.

We explore physical properties of the gas to identify changes that indicate where the line-of-sight is dominated by the ISM versus the outflow. Using our [\ion{O}{III}]~$\lambda$5007 and [\ion{O}{II}]~$\lambda\lambda$3727,3729 emission line maps (see section~\ref{sec:apen_emission_maps}) we construct an $O_{32} = [\ion{O}{III}]/[\ion{O}{II}]$ map as a proxy for ionization parameter. We also determine the gas velocity dispersion ($\sigma_{\rm gas}$) from the width of the [\ion{O}{III}]~$\lambda$5007 emission line, since it is the line with the highest S/N. The instrumental dispersion of our data is $\sim$45~km~s$^{-1}$. The top panel of Fig.~\ref{fig:O32_profile} shows the O32 profile where z is the distance to the midplane of the disk, as defined by the galaxy center. The z=0 point was defined as the highest brightness in IRAC 3.6$\mu$m observations. The blue points and line show the mean value in slices, parallel to the galaxy midplane, of our maps and the shaded area shows the 1$\sigma$ dispersion. At the bottom panel of Fig.~\ref{fig:O32_profile} we show the velocity dispersion as a function of distance from the midplane. The green vertical dashed lines show $r_{90,3.6\mu{\rm m}}$ and $r_{90, g}$. 

The O32 profile in Fig.~\ref{fig:O32_profile} shows that high ionization gas is coincident with the young stars (z $< r_{90, g} = 250$~pc). This is expected because HII regions are located near the galaxy center, causing a higher ionization. The velocity dispersion profile shows a peak at z$\approx$180~pc where SSC~B is located, which coincides with a local minimum in ionization. For z between $r_{90, g}$ and $r_{90,3.6\mu{\rm m}}$ we observe a smooth decrease in ionization and a dip in $\sigma_{\rm gas}$. The decrease in ionization is expected because of the larger distance to ionization sources. For distances larger than $r_{90,3.6\mu{\rm m}}$ we observe an almost flat O32 profile, which is an indication that we are sampling far enough from the disk that stars in the disk are not ionizing the gas. If we do not take into account the maximum due to SSC~B at 180~pc we observe the highest $\sigma_{\rm gas}$ in this region. We expect high velocity dispersion in outflows.

Motivated by the difference in the physical properties that we observe, we define three different zones: (1) The high-ionization zone at ${\rm z}\leq250$~pc is dominated by young stars and subsequently ionized ISM; (2) the decreasing-ionization zone at 250 pc~$<$~z~$<$ 400 pc has a mix of young and old stars, resulting in a smoothly decreasing O32 and dip in velocity dispersion; and (3) the flat ionization zone at z~$>$~400 pc has a much lower surface density of stellar emission (as indicated by the KCWI continuum), producing a flattened low level of O32 and high-velocity dispersion driven by outflows. We refer to these regions as ``Disk dominated'', ``Transition'' and ``Outflow dominated'' respectively.


\section{Measurement of Electron temperature and Metallicity}\label{sec:metallicity}

\begin{figure*}
    \centering
    \includegraphics[width=1\textwidth]{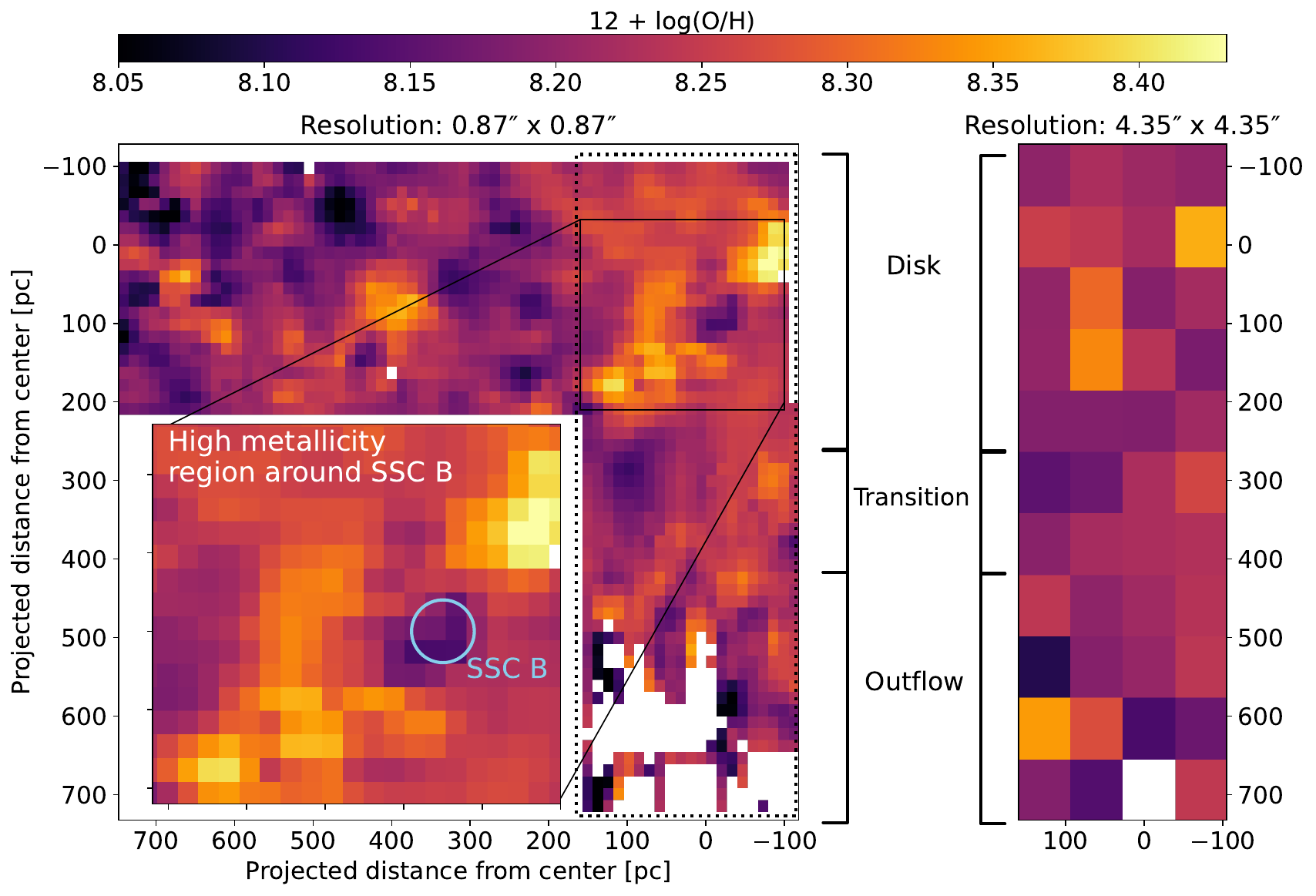}
    \caption{Direct-method metallicity map in 12 + log(O/H) units. On the left we show our entire field of view with a spaxel size of 0.87$\arcsec$ $\times$ 0.87$\arcsec$. In the lower left corner we show a zoom in of a 17$\arcsec$ $\times$ 15$\arcsec$ (260 $\times$ 230~pc) region around SSC~B. On the right we show a map of the metallicity, measured from binned data, in the region marked with a black dotted line in the full resolution map. The spaxel size in the binned map is 4.35$\arcsec$ $\times$ 4.35$\arcsec$. We labeled the disk, transition and outflow regions according to the definition from section \ref{sec:Disk_outflow}}
    \label{fig:Te_Z_maps}
\end{figure*}

With the detection of the strong emission lines plus the detection of [\ion{O}{iii}]~$\lambda$4363 in both the disk and the outflow, we can measure the direct-method oxygen abundance or metallicity. We measure the total O/H abundance by measuring the ionic oxygen abundances in two ionization zones, O$^{+}$ and O$^{++}$ (assuming the contribution of higher ionization levels is negligible), and adding them together. To determine the O$^{++}$/H$^+$ abundance, we use the [\ion{O}{iii}]~$\lambda5007$/H$\beta$ flux ratio with $T_{e}([\ion{O}{iii}])$ calculated from the electron temperature-sensitive line ratio of [\ion{O}{iii}]~$\lambda$4363/$\lambda$5007. $T_{e}$ has a dependence on electron density ($n_{e}$) but for our environment we can assume a low $n_e \approx 10^2$~cm$^{-3}$, where $T_e$ is insensitive to $n_e$ \citep{Izotov2006, Osterbrock2006}. We use the \texttt{getTemDen} function in \texttt{PyNeb} \citep{Luridiana2015}, with the default atomic data and assumed density, to determine $T_{e}([\ion{O}{iii}])$. 

For the O$^+$ abundance, we use the [\ion{O}{ii}]~$\lambda3727$/H$\beta$ flux ratio with the [\ion{O}{ii}] temperature. Our observations do not cover any low-ionization temperature-sensitive emission lines. We therefore assume a simple positive correlation from \cite{Campbell1986}.

\begin{equation}
    \label{eq:TeOIII_TeOII}
    T_e([\ion{O}{ii}]) = 0.7\times T_e([\ion{O}{iii}]) + 3000.
\end{equation}

We measure a mean $T_e([\ion{O}{iii}])$ value of 11,408~K with a standard deviation of 515~K in our entire field of view and a mean $T_e([\ion{O}{ii}])$ of 10,986~K with standard deviation of 360~K. We note that this correlation is based on observations of HII regions and that there is a large scatter \citep{Rogers2021}. 
Future work with broader wavelength coverage is needed to determine if this relationship holds in outflows.

We determine the values for O$^{+}$ and O$^{++}$ relative abundances over $H^{+}$  using the \texttt{pyneb} routine \texttt{getIonAbundance} which uses as input the flux ratio of a given ion over H$\beta$. We use [\ion{O}{III}]~$\lambda$5007 and [\ion{O}{II}$]~\lambda\lambda$3227,9 and the $T_e$. We measure a median metallicity of 12 + log(O/H) = 8.22 in our entire field of view with standard deviation of 0.14. 

The left panel of Fig.~\ref{fig:Te_Z_maps} shows the 12 + log(O/H) map for our KCWI field. We observe some metallicity fluctuation in the disk. In particular we observe higher metallicity around SSC~B. We show a zoom in on this particular region in Fig.~\ref{fig:Te_Z_maps}. The right panel shows the 12 + log(O/H) map from the 4.35\arcsec $\times$ 4.35\arcsec spatially binned data in the region marked by a dotted black line on the left map. The binned data allows us to get a higher SNR on the [\ion{O}{III}]~$\lambda$4363 emission line at large distances from the midplane. In the next sections we will discuss how the SNR affects metallicity measurements and the metallicity fluctuations we observe in the disk and the outflow. 

The presence of [\ion{Fe}{II}]~$\lambda$4360 in some spaxels adds a source of uncertainty to the [\ion{O}{III}]~$\lambda$4363 flux, which can affect our metallicity measurements. The lowest [\ion{Fe}{II}]~$\lambda$4360 to [\ion{O}{III}]~$\lambda$4363 ratio we determine is $\sim$0.1. We might be missing some fainter [\ion{Fe}{II}] emission. If we add 10\% of the [\ion{O}{III}]~$\lambda$4363 flux to itself when computing the metallicity, the median metallicity value changes from 8.22 to 8.17. Thus fainter [\ion{Fe}{II}] emission introduces an uncertainty of up to $\sim$0.05 to our metallicity measurements. On the other hand, the highest [\ion{Fe}{II}]~$\lambda$4360 to [\ion{O}{III}]~$\lambda$4363 ratio we determine is $\sim$0.6. If we add this percent to the [\ion{O}{III}] flux when calculating the metallicities we measure a metallicity of 8.00. With lower spectral resolution data, the [\ion{Fe}{II}] might get completely blended with the [\ion{O}{II}] lines, resulting in an under estimation of the metallicity by up to 0.22.

\section{Electron temperature distribution in the disk}\label{sec:fluctuations}


Here we consider the spaxel-to-spaxel variation of $T_e$, and thus metallicity, in the disk of NGC~1569. We will consider the outflow in a subsequent section. Our data allows us to study the fluctuations of $T_e$ in the ISM on $\sim$15~pc scales and compare them to the global measurement. Differences in emission of [\ion{O}{III}]~$\lambda$4363 and $\lambda$5007 may arise in global measurements \citep{Cameron2023}. Moreover, distinct physical regions may impact a global $T_e$. Our analysis is useful to understand the distribution of metallicity within the disk, and also systematic uncertainties associated with using the direct method to measure metallicity in lower-resolution data \citep[as described in][]{Cameron2023}. 

\begin{figure}
    \centering
    \includegraphics[width=\linewidth]{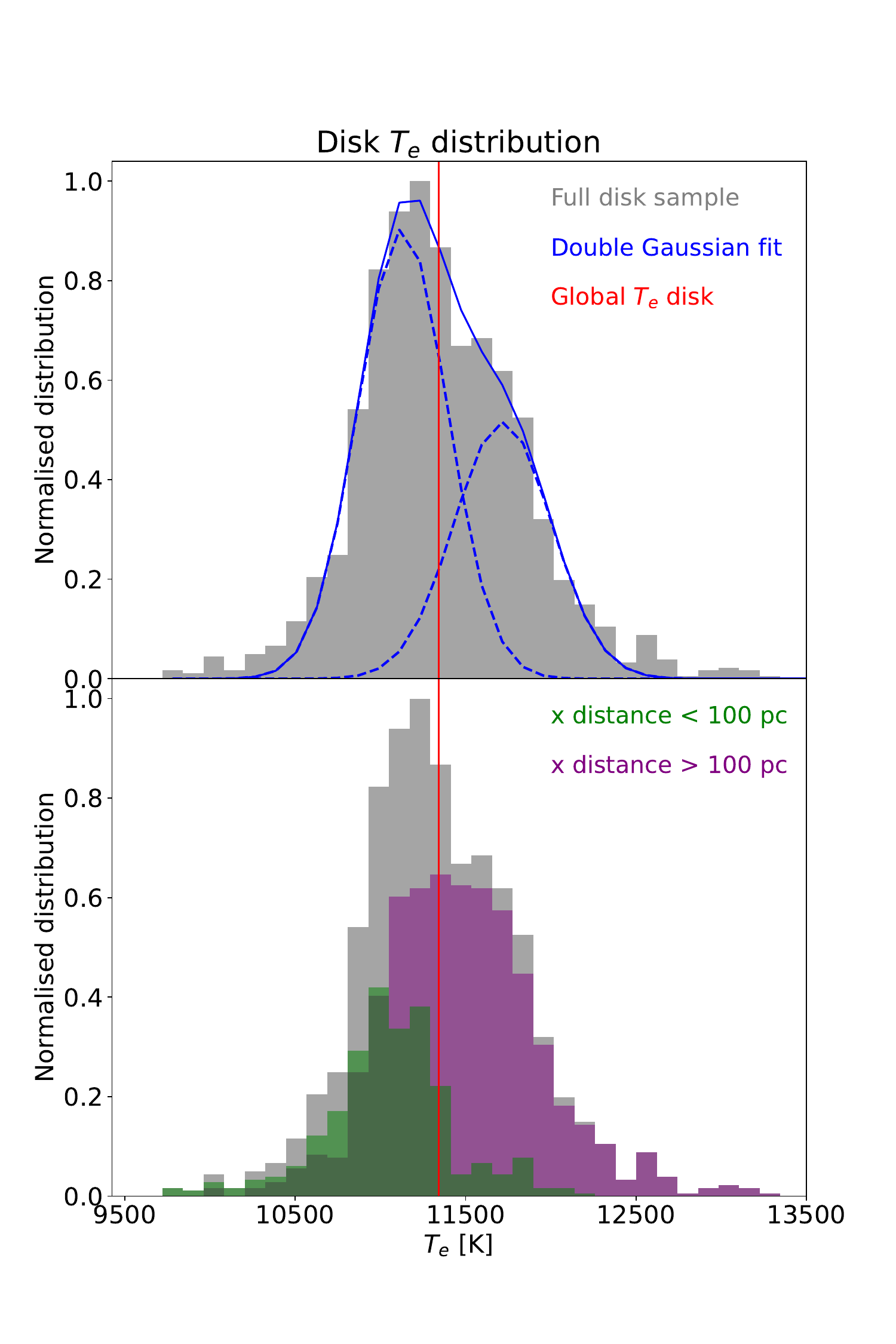}
    \caption{Top panel: The grey histogram shows the distribution of $T_e$ for all spaxels in the 0.87\arcsec $\times$ 0.87\arcsec data in the disk, normalized by the peak of the distribution. We fitted a double Gaussian to the distribution. The blue lines show the best fit. The red vertical line shows the global $T_e$ value, measured from the ratio of the sum of the [\ion{O}{III}]~$\lambda$4363 flux to the sum of the [\ion{O}{III}]~$\lambda$5007 flux of all spaxels in the disk. Bottom panel: Same as top panel, in grey we show the normalised distribution of $T_e$ for all spaxels in the disk in the full resolution data. The green histogram shows the distribution for spaxels within $\pm$100~pc of the galaxy center. In purple we show the histogram for spaxels further than 100~pc from the galaxy center.}
    \label{fig:Te_dist_ISM}
\end{figure}

Fig.~\ref{fig:Te_dist_ISM} shows in grey the electron temperature distribution of spaxels in the 0.87\arcsec $\times$ 0.87\arcsec data in the galaxy disk, normalized by the peak of the distribution. In this section we only consider gas below the ``disk'' region as defined in Section~\ref{sec:Disk_outflow}. The sample has a mean $T_e$ value of $\sim$11420 K with a standard deviation of $\sim$483 K. We measured the global $T_e$ of the disk from the summed spectra of all spaxels in the disk. We obtained a value of $\sim$11409 K. The global measurement for $T_e$ is consistent with our sample's mean value of all spaxels in the ISM. This implies that $T_e$ fluctuations in scales of around $\sim15$~pc do not affect the global measurements of metallicities using the direct method. We note that our observations do not cover the entire disk. Wider field observations may still generate a discrepancy.

We do however observe an asymmetric profile in Fig.~\ref{fig:Te_dist_ISM}. There is a feature at $\sim$12000~K, which may indicate a secondary component to the distribution of $T_e$ in NGC~1569.  We, therefore, tested this by fitting a double Gaussian model to the distribution, and compare this to a single Gaussian fit to the entire distribution. We use a BIC comparison of the models, and find $\Delta$BIC $\sim$2200. The double Gaussian model is, therefore, strongly favored. 

The top panel of Fig.~\ref{fig:Te_dist_ISM} shows the best fit for the distribution of $T_e$ values in the ISM. The two Gaussian components have centroid values of 11154~K and 11750~K. In the bottom panel of Fig.~\ref{fig:Te_dist_ISM} we show that this multi-component behavior may be recovered by the position within the galaxy. In gray we show again the full distribution in the ISM, but now we also show two histograms based on location within NGC~1569. The green histogram represents spaxels in the ISM that are within $\pm$100~pc of the galaxy center. In purple we show the distribution of spaxels in the ISM that are further than 100~pc from the galaxy center. Those spaxels in the galaxy center have a median $T_e$ value of 11103~K,  while those at larger distance have median $T_e$ of 11507~K. This implies that gas at larger distances from the center of the galaxy, in the disk zone,  tend to have higher $T_e$ values or lower metallicities.

While metal-poor gas at larger distances is often explained by accretion \cite[e.g.][]{Bresolin2012,Cameron2021}, the center of NGC~1569 has multiple structures that may impact the local metal content. We note that signatures of accretion have been observed in NGC~1569 in atomic hydrogen \citep{Muhle2005}. The HI accretion is, however, at significantly larger distance. We will therefore consider if the bright SSC in NGC~1569 is driving these local changes in $T_e$.  



In Fig.~\ref{fig:Te_Z_maps} SSC~B is located at the position [0, 150~pc]. At the bottom left of the plot we show a zoom in to the region around SSC~B. In and around SSC~B we observe a complex metallicity substructure. In those spaxels corresponding to the location of the cluster, we observe a low metallicity. Immediately outside SSC-B, but less than a distance of $\pm$200~pc we identify two regions with elevated metallicity compared to both the disk and outflow.  The first is located roughly south of the SSC, the second one is located to the north of SSC~B. From the emission line maps (Fig.~\ref{fig:emissionmaps}) and the HST image we observe that the region surrounding SSC~B has low flux. This region of high metallicity also corresponds to a region of low ionization, which can be observed in the dip at z $\approx$ 100~pc in the log(O32) profile (Fig.~\ref{fig:O32_profile}). The lower value of O32 could happen if the [\ion{O}{III}] emission was truly due to lower mass density, and thus generating less collisional excitation. Moreover, inspection of HI maps in \cite{Johnson2013} for the same region likewise show a decrease in HI flux. Previous authors argue that supernovae from SSC~B produces a superbubble \citep{Westmoquette2007}. The lack of gas and metal enrichment seems consistent with this picture.

From the [\ion{O}{II}]~$\lambda\lambda$3726,9 doublet we measure a median electron density ($n_e$) of the gas in the ISM of 42~cm$^{-3}$. Our [\ion{O}{II}]~$\lambda$3729/$\lambda$3726 line ratios are sufficiently low to consider this $n_e$ near to an upper-limit. The high metallicity region around SSC~B, shown in the zoom-in panel in Fig.~\ref{fig:Te_Z_maps}, has on average a higher $n_e$. The southern region has a particularly high $n_e$ of $\sim1300$~cm$^{-3}$, 30 times higher than the mean ISM value. This region has a median $T_e([\ion{O}{iii}])$ value of $\sim$10870~K, lower than the ISM median value of 11430~K. A map of the measured $T_e$ and $n_e$ is shown in Fig.~\ref{fig:Te_ne_map}. The ionized gas radiation pressure ($P_{\rm HII}$) is proportional to the product of $T_e$ and $n_e$. The mean value for $P_{\rm HII}/k_B$ in the southern high metallicity region is 2.2$\times 10^{7}$~(cm$^{-3}$~K). The median value of $P_{\rm HII}/k_B$ in the disk is 9.7$\times 10^{5}$~(cm$^{-3}$~K). The median $P_{\rm HII}$ in the disk is consistent with values for HII regions in NGC~300 \citep{McLeod2021}. The value in the southern high metallicity region is $\sim$1.5 orders of magnitude higher than their values. 

NGC~1569 had a starburst phase 10--20 Myrs ago \citep{Israel1988, Hunter2000} where SSC~B was formed \citep{Hunter2000} with a mass of 2.3 $\times 10^5 M_{\odot}$ \citep{Gilbert2002}. This cluster probably produced hundreds of supernovae that resulted in the superbubble we observe \citep{Sanchez-Cruces2015}. The metal-rich gas surrounding the SSC could be expelled directly from the high star formation in SSC~B into the surrounding ISM. A simple explanation is that the gas surrounding the SSC~B more heavily favors SNe ejecta, rather than the pre-existing ISM, which allows it to be more metal rich. 


The physical properties measured around SSC~B could be used to identify starburst (or post-starburst) regions in other galaxies, which may indicate areas in which feedback has disrupted star formation and possibly are the sites of recently disrupted molecular clouds. We list the properties of this starburst region in NGC~1569 in comparison to the ISM of the galaxy. We note that in other galaxies the exact values may change, nevertheless these may provide guidance. The properties of the region include:

\begin{addmargin}[1em]{2em}
  \noindent $\bullet$ High electron density is among the most distinct differences. In this region $n_e$ reaches values over 10$^3$~cm$^{-3}$, while being near to the low density limit for [OII] in the rest of the disk. \\
 \noindent $\bullet$ A decrease in surface brightness of Balmer lines, with respect to the surrounding galaxy, by a factor of 5-10$\times$.\\
 \noindent  $\bullet$ Lower electron temperature (high metallicity) than the disk. In NGC~1569 this is of order $\Delta T_e \sim 1000$~K.\\
 \noindent  $\bullet$ Lower ionization in the cavity. In NGC~1569 this is of order a factor of 2-3 $\times$ higher O32 in the disk. We note that NGC~1569 is a higher ionization system. The exact difference may vary. 
\end{addmargin}

\section{Outflow Enrichment}
\label{sec:enrichment}

As we have shown in Fig.~\ref{fig:Te_Z_maps}, we measure the metallicity gradient along the minor axis (z) in the image plane, at a distance of $\sim$700~pc with a resolution of 15~pc. We consider gas in the outflow region to be at ${\rm z} > 400$~pc, as defined in Section~3.3. We also defined the ``transition'' region for $250 < {\rm z} < 400$~pc, where the gas is probably a mix of ISM and outflow. We note that there is not a hard boundary at which an outflow begins, as gas would begin flowing outward from the source of the driving mechanism. We compare the results in this section for ${\rm z} > 400$~pc and ${\rm z} > 250$~pc and the results do not change substantially. Therefore, in this section, we will include spaxels from the transition region with ${\rm z} > 250$~pc, and when appropriate discuss any changes with z-axis distance from the galaxy.   

\begin{figure}
    \centering
    \includegraphics[width=\linewidth]{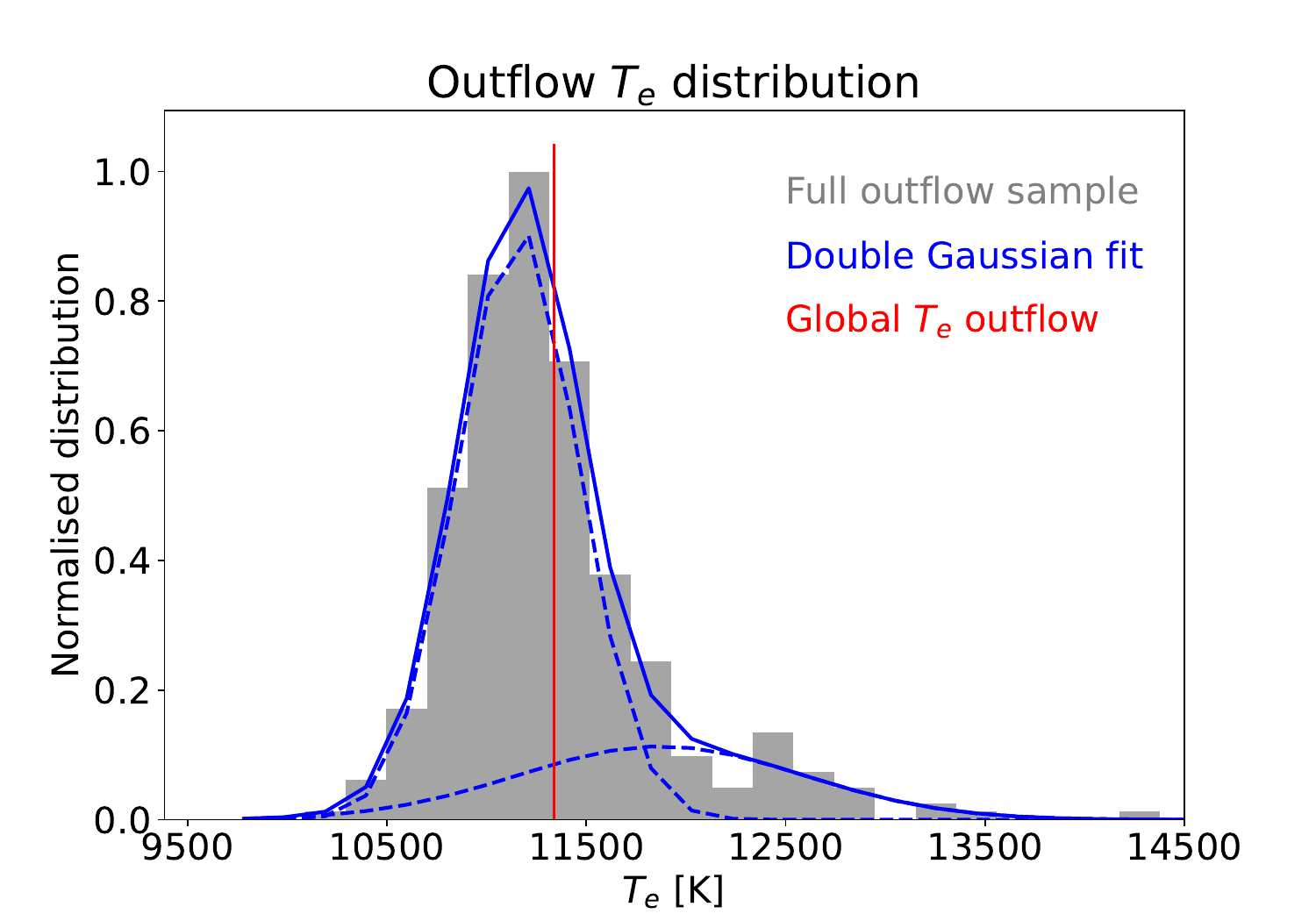}
    \caption{$T_e$ distribution of spaxels in the outflow. Grey histogram shows the distribution of $T_e$ in the 0.87\arcsec $\times$ 0.87\arcsec data, normalized by the peak of the distribution. In blue we show a double Gaussian fit to the distribution. The red vertical line shows the global $T_e$ measured from the sum of the spectra of all spaxels in the outflow.}
    \label{fig:Te_dist_out}
\end{figure}


The grey shaded histogram in Fig.~\ref{fig:Te_dist_out} shows the distribution of $T_e$ for spaxels in the outflow, normalized by the peak of the distribution, for the 0.87$\times$0.87~arcsec$^2$ spaxel data. We observe that the outflow sample shows a strong peak with a median $T_e$ of 11296~K with standard deviation of 630~K. The distribution shows a tail towards large $T_e\sim 12500-14000$~K. As for the disk, we fit single and double Gaussian models to the distribution of $T_e$ in the outflow. We found $\Delta {\rm BIC} \sim 100$. This suggests that the double Gaussian model is preferred. 

We identified all spaxels with $T_e$~>~12500~K to have a mean SNR on the [\ion{O}{III}]~$\lambda$4363 line of 3.8, while spaxels with $T_e$~<~12500~K have a mean SNR of 6.8. Likewise, spaxels with high electron density are the faintest detections of emission lines in our maps. To investigate the impact of low SNR on estimation of $T_e$, we binned our the data cube to obtain spaxels with size 4.3\arcsec~$\times$~4.3\arcsec. In Fig.~\ref{fig:Te_Z_maps} we show a metallicity map of the binned data in a region that includes the outflow region and the disk region directly above the outflow. We observe that at the lower spatial resolution we are able to measure the metallicity towards larger distances from the midplane. This is because of the increase in SNR on the [\ion{O}{III}]~$\lambda$4363 auroral line. 

\begin{figure}
    \centering
    \includegraphics[width=\linewidth]{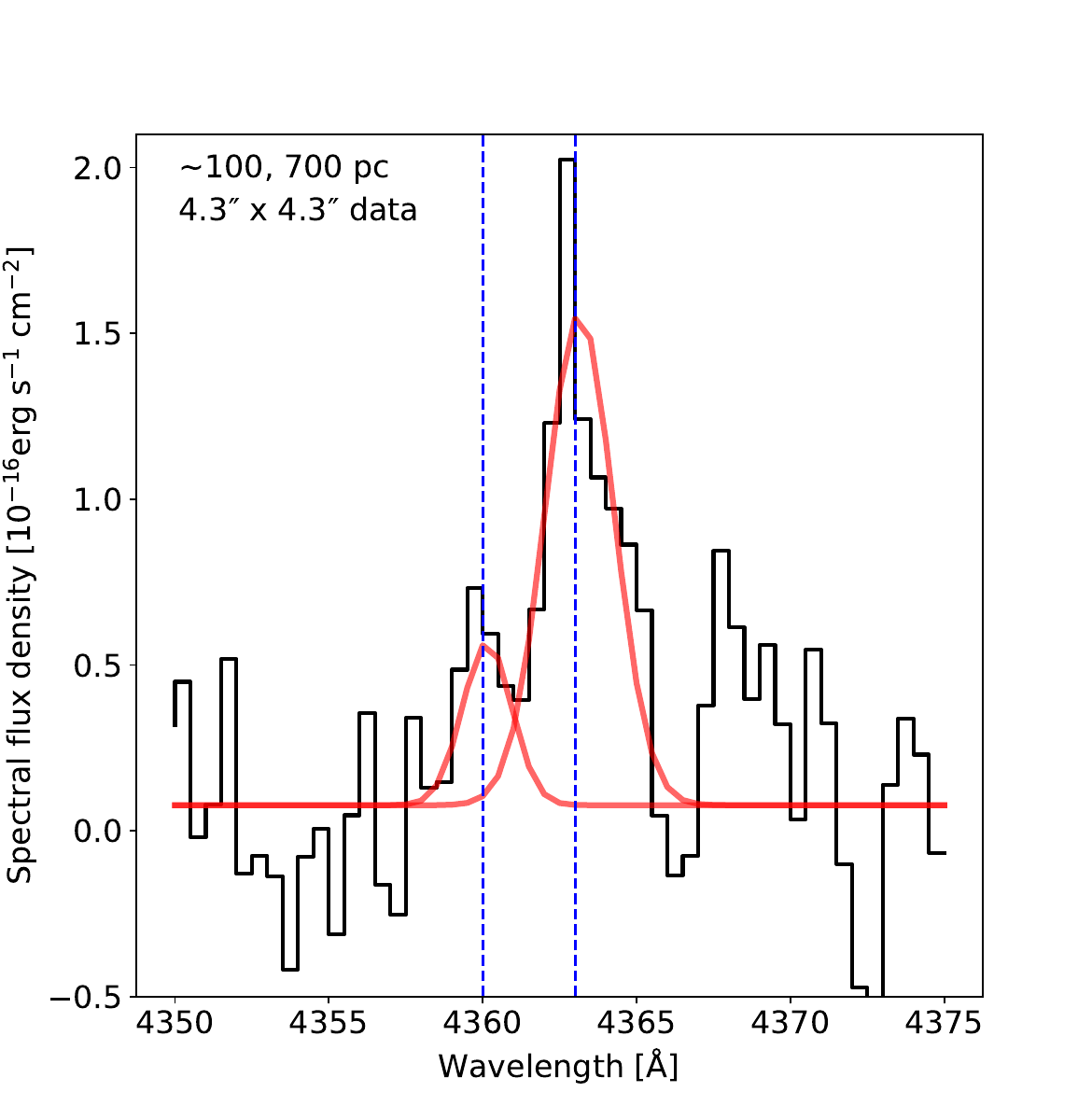}
    \caption{ Spectrum of the binned 4.3\arcsec $\times$ 4.3\arcsec data at the position $\sim$100, 700~pc. The black line shows the data in a wavelength region around the [\ion{O}{III}]~$\lambda$4363 emission line. The red line shows the best fit. The second Gaussian at 4360~{\AA} corresponds to the [\ion{Fe}{II}] line. This [\ion{Fe}{II}]~$\lambda$4360 feature is not detected in the higher spatial resolution data (0.87\arcsec $\times$ 0.87\arcsec) in the same region. This results in a higher flux estimation and thus a higher $T_e$ measurement for the higher resolution data.}
    \label{fig:binned_spec}
\end{figure}

Fig.~\ref{fig:binned_spec} shows a spectrum of the [\ion{O}{III}]~$\lambda$4363 emission line in a binned spaxel at the position $\sim$100, 700~pc. We estimate a SNR of $\sim$23 for the [\ion{O}{III}]~$\lambda$4363 line. From Fig.~\ref{fig:emissionmaps} we observe that this region in the high resolution map does not show many [\ion{O}{III}]~$\lambda$4363 detections. The spaxels where we do find detections have a mean SNR of 3.3. We measure a mean metallicity of 12+log(O/H)=8.13 and a mean $T_e$ of 12834~K for the unbinned data. In the lower resolution map we observe that this region corresponds to a metallicity of $\approx$8.18 and a $T_e$ of $\approx$12000~K. In their simulations, \cite{Cameron2023} found a systematic bias where the $T_e$ measured from [\ion{O}{III}]~$\lambda$4363/[\ion{O}{III}]~$\lambda$5007 overpredicts the true $T_e$ by over $\sim1000$~K for gas of $T_e$=12000K. Our offset is high, though not as high. 

The red line in Fig.~\ref{fig:binned_spec} shows the best fit model for the binned data at $\sim$100, 700~pc using the method described in section~\ref{sec:linefits}. For the spaxel at this position, the BIC method selects a double Gaussian fit in the binned data. The fainter emission bluewards of the [\ion{O}{III}]~$\lambda$4363 line corresponds to the [\ion{Fe}{II}]~$\lambda$4360 line \citep{Andrews2013, Curti2017}. The [\ion{Fe}{II}] peak is significantly lower than the peak of  [\ion{O}{III}]~$\lambda$4363, and is near to the SNR limit in the binned spectrum. In the unbinned spectrum, identifying [\ion{Fe}{II}] will likely require high signal-to-noise, which may not be available in the data. If \texttt{threadcount} fits a single Gaussian to both the [\ion{Fe}{II}]  and [\ion{O}{III}] features, this would artificially increase the flux attributed to [\ion{O}{III}]~$\lambda$4363. This, then, would result in an overestimation of $T_e$. 
We therefore caution the reader that the high $T_e$ or low metallicity could be the result of unaccounted for [\ion{Fe}{II}] contamination.

The possible contamination of [\ion{Fe}{II}] in the outflow motivates us to investigate this further. The [\ion{Fe}{II}]~$\lambda$4360 feature is, however, blended with [\ion{O}{III}]~$\lambda$4363, which complicates measurement. \cite{Curti2017} showed that the [\ion{Fe}{II}]~$\lambda$4288 correlates with the presence of [\ion{Fe}{II}] $\lambda$4360. We estimate [\ion{Fe}{II}]/H$\beta$ in all of our spaxels. We find that in the disk, the median value is  0.005, while in the outflow the ratio is twice as large,  [\ion{Fe}{II}]/H$\beta\sim$0.009.  We note that this provides an alternate view of enrichment, in which there is stronger [\ion{Fe}{II}] in the wind than in the disk. Indeed, \cite{Curti2017} showed that  [\ion{Fe}{II}] emission is correlated with metallicity. The [\ion{Fe}{II}] flux is, however, more difficult to interpret. This may be heavily affected by ionization state. More work is needed to fully interpret these enrichment features. Nevertheless, we take it that this further supports our interpretation that the high $T_e$ tail is likely due to [\ion{Fe}{II}] contamination. This should be considered for all $T_e$ measurements that rely on low SNR measurements of [\ion{O}{III}]~$\lambda$4363. 




If we assume that the high $T_e$ tail is an artificial effect, then there is not a clear bimodality in $T_e$ in the outflow. This is unlike the disk, which we have discussed shows a variation that may indicate different physical areas of the galaxy. In Fig.~\ref{fig:emissionmaps}, we show that in both H$\beta$ and [\ion{O}{III}] emission the outflow contains substructure. Beginning at 300~pc from the galaxy center, and extending across our field-of-view there is a bright filament in both emission lines. This is likewise seen in the HST image (Fig.~\ref{fig:pointings}) and has been associated with outflowing gas \citep{Westmosquette2008,Johnson2013}. This change in gas substructure does not appear to impact the local metallicity. Fig.~\ref{fig:Te_Z_maps} shows the metallicity map where the metallicity does not follow the filamentary structure we observe in emission line fluxes. We do not find any significant correlation between the metallicity of the gas and emission line fluxes or ionization in the outflow. This may imply that small scale fluctuations in the $T_e$ of outflows are not a prominent effect.


\section{Low Electron Density in the Outflow of NGC~1569}\label{sec:density}

In order to estimate the mass of ionised gas in an outflow, a value for the electron density is necessary. As in the disk, our data allows us to estimate the electron density of the outflow of NGC~1569. The [\ion{O}{II}] ratio can be used to measure $n_e$ for values of $0.347 < [\ion{O}{II}] \lambda 3729 / \lambda 3726  < 1.5$ \citep{Pradhan2006}, where the range is bounded by the so-called low and high density limits. Outside this range the ratio reaches asymptotic values. 

We detected the [\ion{O}{II}] doublet across our entire field-of-view and calculated $n_e$. Within the measurements errors in emission line fluxes, in the region we defined as the outflow, 43\% of the spaxels have values of $[\ion{O}{II}] \lambda 3729 / \lambda 3726  > 1.5$. This indicates that $n_e$ is below the low density limit. The lowest $n_e$ we measure, due to measurements uncertainties, is 26~cm$^{-3}$. The fraction of spaxels in the low density limit increases with distance to the galaxy center, suggesting that $n_e$ in the outflow decreases with distance from the midplane. For spaxels above the low density limit the median $n_e$ is 82~cm$^{-3}$, this is an upper limit on the outflow electron density. 

Typical assumed values for $n_e$ in outflows range from $200 - 500$~cm$^{-3}$. These values are  derived from large samples of galaxies, mostly high-metallicity ULIRGS and starbursts, with broad emission lines \citep{Forster2019, Fluetsch2021}. 

Resolved $n_e$ measurements have been done for a small sample of outflows. \cite{Xu2023} measured $n_e$ for M~82, finding 200~cm$^{-3}$, which decays as $n_e \propto z^{-1.17}$. \cite{Bik2018} measured $n_e$ from [\ion{S}{II}] in the outflow of ESO 0338-IG04, a galaxy with similar metal content to NGC~1569. They found a profile that decays quite steeply with values $10-100$~cm$^{-3}$. Our measurement of $n_e$ in the outflow of NGC~1569 is consistent with the low $n_e$ at large distances. 

We note the implications for studies that aim to estimate the mass-loading factor of winds, especially those at larger redshift. Our result and that of \cite{Bik2018} are consistent with a scenario in which outflows from lower metallicity galaxies have $n_e \sim 10-100$~cm$^{-3}$, which is lower than commonly adopted values. The impact of decreasing $n_e$ is to increase the mass loading factor. Recently, \cite{Carniani2023} measured the mass outflow rate for galaxies at redshifts $3<z<9$ and found a mass loading factor that is $\sim$100 times higher than those in $z=0$ galaxies. However, it is almost certainly the case that metallicity evolves significantly across this redshift range. More work on the systematics of $n_e$ in outflows for galaxy properties is needed in order to improve estimation from large surveys and accurately inform galaxy evolution modelling.

\section{Metal Loading in NGC~1569}\label{sec:loading}

\begin{figure*}
    \centering
    \includegraphics[width=\linewidth]{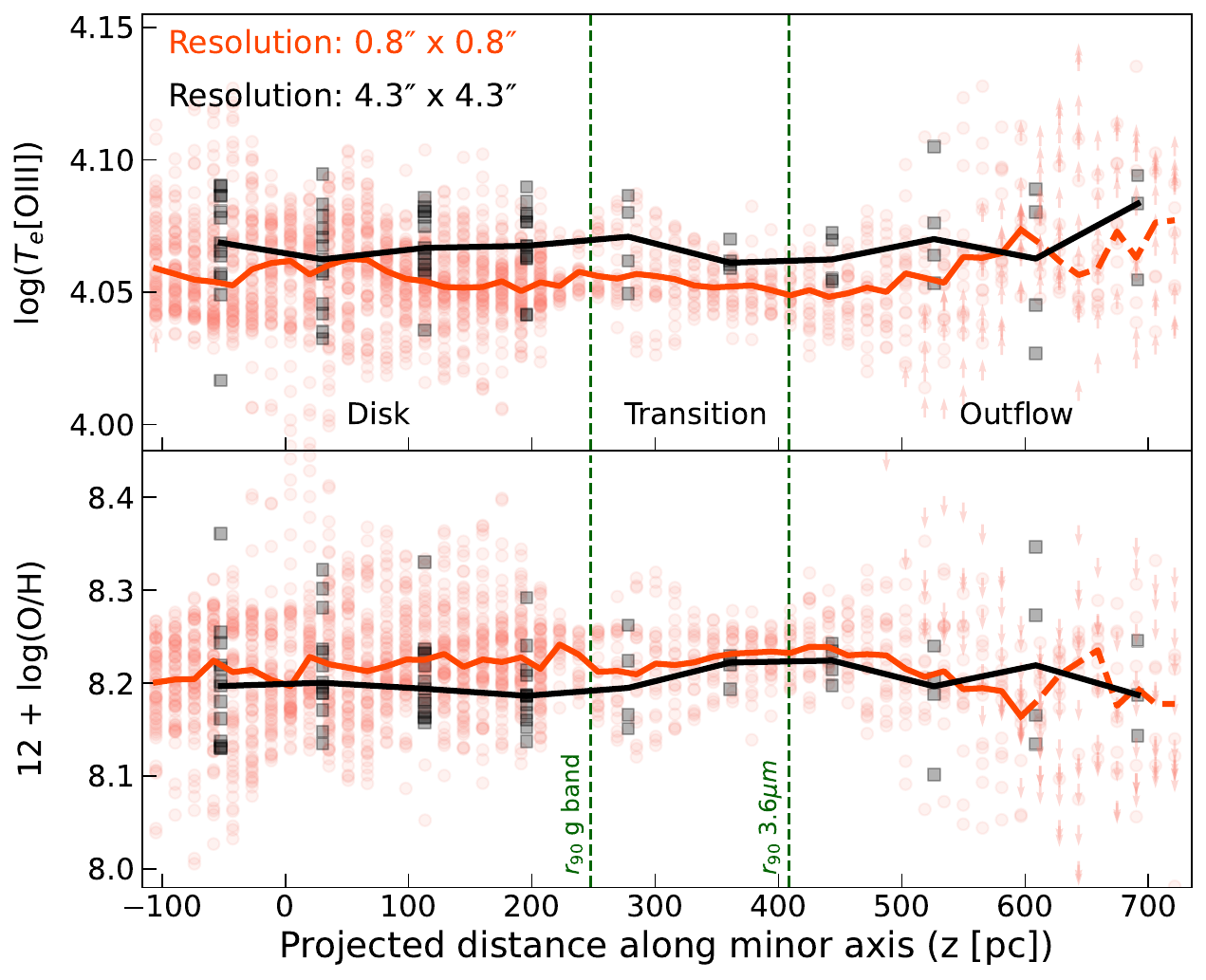}
    \caption{The z-axis profile of $T_e$ and metallicity of NGC~1569. The regions, defined previously, are denoted with vertical dashed lines and labels above the bottom panel. Orange points indicate the individual $\sim$15~pc spaxels and orange line indicate upper limits. The solid orange line indicates the median in bins of each row perpendicular to the z-axis, where upper limits are less than 40\% of the total measurements. The dotted orange line shows the median in bins where upper limits are more than 40\% of the data, in this cases upper limits were ignored in the median calculation. The black squares represent the metallicity measured in binned spaxels with size $\sim$65~pc. The metallicity profile in NGC~1569 is consistent with a flat distribution, with very similar metallicity in the disk as in the wind.}
    \label{fig:Te_Z_profile}
\end{figure*}

Fig.~\ref{fig:Te_Z_profile} shows the electron temperature and metallicity profiles along the minor axis of NGC~1569 for the high resolution (0.8\arcsec $\times$ 0.8\arcsec spaxels) and lower resolution data (4.3\arcsec $\times$ 4.3\arcsec). Within the measurement uncertainty, our observations are consistent with a flat metallicity gradient between the disk and outflow. The median metallicity in the disk region for the high resolution data is 8.21 with a standard deviation of 0.06 and the median metallicity in the outflow region is 8.22 with a standard deviation of 0.06.  For the lower resolution data the median metallicity in the disk is 8.19 with a standard deviation of 0.05 and 8.20 in the outflow with standard deviation of 0.06. In both the disk and the outflow the binned metallicity profile is consistent  with the full resolution profile within the scatter of our data. We therefore conclude that the incomplete coverage of the finer spatial resolution does not impact the metallicity gradient along the minor axis of the image.

\cite{Garduno2023} used strong lines to measure the metallicity in NGC~1569. They found a mean metallicity of 12~+~log(O/H)~=~8.12. This is similar to our values, considering the typicaly systematic uncertainties between different methods of measuring metallicity. They also found little variation of metallicity with other properties and distance to the galaxy center, which is consistent with our flat metallicity profile.




Recently, \cite{Vijayan2023} carried out box-simulations of supernovae-driven winds which include metallicity effects on the outflow. They included a range of metallicities in the simulation and showed a result with 0.2~$Z_{\odot}$, which is appropriate for comparison to NGC~1569. The shape of the simulated metallicity profile in \cite{Vijayan2023} depends on the gas phase and the initial metallicity of the galaxy. For the warm ionized gas, the profile will be steeper for lower metallicity galaxies. The profile for galaxies with $Z_{\rm ISM} = 0.2$~Z$_{\odot}$ goes from Z$_{\rm ISM}$ in the midplane to $\sim$1.1~Z$_{\rm ISM}$ at distances of $\sim$1~kpc from the midplane. NGC~1569 has a metallicity of $\sim$0.25~Z$_{\odot}$. Considering error bars in our metallicity profile for NGC~1569 it is consistent with the results in \cite{Vijayan2023}. 

The only other direct-method outflow metallicity profile was observed by \cite{Cameron2021}. They measured the metallicity profile along the minor axis of Mrk~1486 up to $\sim$1.6~kpc away from the midplane. They found that the metallicity at the larger distance was 1.6 times $Z_{\rm ISM}$. Mrk~1486 has metallicity of $\sim0.05-0.1$~Z$_{\odot}$. This is likewise consistent with the simulation of \cite{Vijayan2023}, as it indicates a steeper profile. 

The total mass-loading in the \cite{Vijayan2023} simulation is of order $\dot{M}_{\rm out}/{\rm SFR}\sim 1$. In NGC~1569 \cite{McQuinn2019} estimated the mass-loading of ionised gas to be $\dot{M}_{\rm out}/{\rm SFR}\sim 3$. This does not include the other phases. Results from VLA observations suggest a significant mass component exists with HI gas \citep{Johnson2012}. The total mass loading of NGC~1569 is likely higher. It is not clear how the dependence of mass-loading may, or may not, impact the metallicity profile of the wind. Nevertheless, the current agreement between the very few observations of metal-loading in winds with simulations seems encouraging.

\begin{figure}
    \centering
    \includegraphics[width=\linewidth]{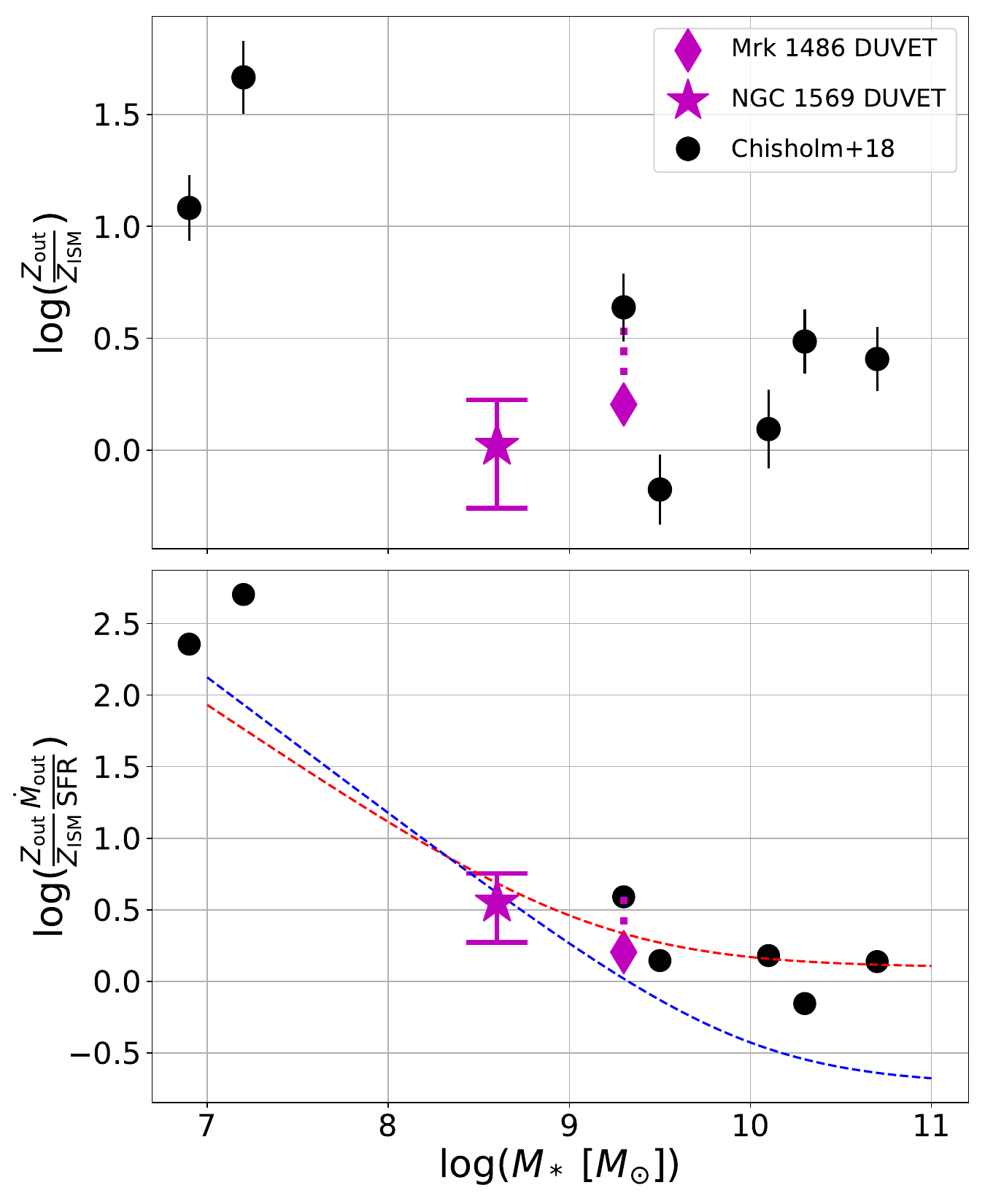}
    \caption{Top panel: $Z_{\rm out}/Z_{\rm ISM}$ as a function of $M_{\star}$. The magenta star shows the result of this work for NGC~1569. The magenta error bars show the two most extreme values depending on what regions we consider as ISM and outflow. The black dots show the results from \protect\cite{Chisholm2018} where they measured $Z_{\rm out}$ from absorption lines and $Z_{\rm ISM}$ using the direct method. The magenta diamond shows the result from \protect\cite{Cameron2021} for Mrk~1486 using the direct method for $Z_{\rm out}$ and $Z_{\rm ISM}$. The magenta diamond is connected to the \protect\cite{Chisholm2018} measurement for the same galaxy with a magenta dotted line. Bottom panel: $(Z_{\rm out}/Z_{\rm ISM}) \times (\dot{M}_{\rm out}/{\rm SFR})$  as a function of $M_{\star}$. Same colors as the top panel. Red and blue lines show analytical predictions in order to reproduce different MZR in the literature. The red line shows the scaling for the MZR in \protect\cite{Tremonti2004} and the blue line shows the scaling for the MZR in \protect\cite{Denicol2002}
    }
    \label{fig:metal_loading}
\end{figure}

\cite{Chisholm2018} measured the metallicity in the outflow of 7 galaxies using absorption lines. They compared the absorption line wind metallicity to direct-method metallicity measurements of the ISM for each corresponding galaxy. The top panel in Fig.~\ref{fig:metal_loading} shows $Z_{\rm out}/Z_{\rm ISM}$ as a function of stellar mass of the galaxy. In black are the \cite{Chisholm2018} results. They found a steep dependence of increasing $Z_{\rm out}/Z_{\rm ISM}$  with decreasing $M_{\star}$. For galaxies with  ${M_{\star} < 10^{7.5}}$~M$_{\odot}$ outflows are more enriched than the ISM by a factor higher than 10$\times$. We note, however, there are only 2 galaxies of this mass and no galaxies in the \cite{Chisholm2018} sample between $M_{\star} = 10^{7.5}-10^{9}$~M$_{\odot}$.

The magenta symbols in Fig.~\ref{fig:metal_loading} indicate those measured using $T_e$-based metallicities in the outflow. Given the significant systematics, the measurement from  \cite{Cameron2021} for Mrk~1486 is not inconsistent with that from the absorption lines. For NGC~1569, using the median metallicity in the disk and the median metallicity in the outflow from the full resolution data we get $Z_{\rm out}/Z_{\rm ISM} \approx 0.93$. Using the binned data we measure $Z_{\rm out}/Z_{\rm ISM} \approx 1.05$. We plotted the value for the binned data as a magenta star in Fig.~\ref{fig:metal_loading}. 


The two measurements using direct metallicity method to determine the metallicity in the outflow are within the range of $Z_{\rm out}/Z_{\rm ISM}$ for the high-mass galaxies measured with absorption lines. It is not clear if the $Z_{\rm out}/Z_{\rm ISM}$ for NGC~1569 is deviating from the trend of low mass galaxies having higher $Z_{\rm out}/Z_{\rm ISM}$. NGC~1569 is the third lowest mass galaxy in the figure, and yet has the second lowest $Z_{\rm out}/Z_{\rm ISM}$.  More measurements are needed, in particular for lower mass galaxies.

From the $Z_{\rm out}/Z_{\rm ISM}$ measurements we can calculate the metal loading factor ($\zeta$) using Equation~\ref{eq:metalloading}, we show our results in the bottom panel of Fig.~\ref{fig:metal_loading} as the magenta star. For NGC~1569 we adopt $\frac{\dot{M}_{\rm out}}{\rm SFR}\sim3$ from \cite{McQuinn2019} using H$\alpha$ imaging. The red and blue dotted lines show two models from \cite{Peeples2011} that were calibrated to match the MZR observations of \cite{Tremonti2004} and \cite{Denicol2002}. The steep dependence of metal-loading with galaxy mass is necessary to reproduce the MZR. 

We show in Fig.~\ref{fig:metal_loading} that both $T_e$-based outflow metallicities, the magenta markers, are consistent with the $M_{\star}$ - $(Z_{\rm out}/Z_{\rm ISM}) \times (\dot{M}_{\rm out}/{\rm SFR})$ relationship found by \cite{Chisholm2018} using absorption lines. We do, however, urge significant caution for these results. This agreement depends strongly on the mass-loading of the galaxy; in fact the mass loading is a larger fraction of the metal-loading than $Z_{\rm out}/Z_{\rm ISM}$. There is a very significant amount of uncertainty in measuring mass-loading of galaxies. Most notably, this mass-loading is only a single phase, which we know to be a smaller fraction of the outflow than cold phase gas. $Z_{\rm out}/Z_{\rm ISM}$, which is a much more robust measurement when using auroral lines than mass-loading,  remains low for its mass. More observations are needed, especially at low stellar mass, to determine if this trend continues.

We observe spatial variations of metallicity in the disk and the outflow regions as shown by the metallicity map, Fig.~\ref{fig:Te_Z_maps}. For example, the region around SSC~B shows a particularly high metallicity compared to the disk. To account for this spatial variation, we calculate the minimum $Z_{\rm out}/Z_{\rm ISM}$ using the largest metallicity measured in the binned data in the disk and the lowest metallicity measured in the binned data in the outflow. We obtain a minimum $Z_{\rm out}/Z_{\rm ISM}$ = 0.55. Likewise, we measure the maximum $Z_{\rm out}/Z_{\rm ISM}$ using the lowest metallicity measured in the disk and the highest metallicity measured in the outflow. We obtain a maximum $Z_{\rm out}/Z_{\rm ISM}$ = 1.6. We plot the maximum and minimum $Z_{\rm out}/Z_{\rm ISM}$ and $(Z_{\rm out}/Z_{\rm ISM}) \times (\dot{M}_{\rm out}/{\rm SFR})$ as error bars to the magenta star in Fig.~\ref{fig:metal_loading}. Even considering the minimum and maximum values, NGC~1569 shows similar metal-loading to high-mass galaxies. We consider that the median values for $Z_{\rm ISM}$ and $Z_{\rm out}$ are a better comparison to \cite{Chisholm2018} measurements because they come closer to a metallicity of the whole disk and the whole outflow.

\section{Summary}\label{sec:summary}

In this paper we use observations from Keck/KCWI to study the metallicity in the disk and wind of nearby, starbursting dwarf galaxy NGC~1569. Our observations are the second-ever measurement of the metal loading in a galaxy using the direct method metallicity for both the wind and the outflow. The metal loading factor is a necessary input for simulations of galaxy evolution \citep[e.g.][]{Peroux2020,Wright2021}, as well as explanations of the MZR and metallicity gradients \citep{Peeples2011,Sharda2021}.  We detect the electron temperature sensitive [\ion{O}{III}]~$\lambda$4363 emission line across nearly our entire field of view. We measure $T_e$ from the [\ion{O}{III}]~$\lambda$4363/$\lambda$5007 ratio, then measure the metallicity using the direct method and find $12+\log({\rm O/H)}\sim8.2$ in both disk and outflow.

Our main result is that the metallicity of the wind is comparable to the disk ($Z_{\rm out}/Z_{\rm ISM}\sim 1$). We find that direct-method based metal loading factors are consistent with the closed-box derivations of the dependence of metal-loading factors (${\zeta = (Z_{\rm out}/Z_{\rm ISM}) \times (\dot{M}_{\rm out}/{\rm SFR})}$) on galaxy stellar mass, and are likewise consistent with previous results from absorption line studies \citep{Chisholm2018}. A trend of mass and $Z_{\rm out}/Z_{\rm ISM}$ is not clear for the current measurements. As this is a more robust observational quantity to estimate, this may be a concern for the metal-loading of galaxies. A larger sample of galaxies with direct metallicity measurements in the disk and outflow are needed in order to see if a clear trend with mass exists. 

The first resolved direct-method measurement of outflow metallicity from \cite{Cameron2021} used 0.7~kpc resolution observations from Keck/KCWI to study the outflow and inflow of Mrk~1486. The observations we present here (Fig.~\ref{fig:pointings}) have a spatial resolution of $\sim$15~pc, much finer than the previous work from \cite{Cameron2021}. We can, therefore, investigate the fluctuation of $T_e$ on scales of $\sim10-100$~pc, and thus determine how these may impact global measurements. We do not observe any significant metallicity variation in the outflow, which suggests that small $T_e$ fluctuations in the outflow are not a prominent effect. 

We show that the sensitivity of the [\ion{O}{III}]~$\lambda$4363 line must be taken into consideration when using it to estimate $T_e$. The reason for this is, as has been shown by \cite{Curti2017}, [\ion{Fe}{II}]~$\lambda$4360 emission can contaminate the [\ion{O}{III}]. At low SNR this may not be easily identified. Very little is known about [\ion{Fe}{II}]~$\lambda$4360. We find that even though our emission line SNR is above ${\rm SNR}\sim5-10$, spatially binning the data reveals the presence of [\ion{Fe}{II}]~$\lambda$4360. The systematic bias of not including [\ion{Fe}{II}]~$\lambda$4360 in a fit is to increase the flux attributed to [\ion{O}{III}]~$\lambda$4363, and thus increase the estimated $T_e$. We therefore re-estimate $T_e$ in spatially binned spectra, which can identify [\ion{Fe}{II}]. We find that this has the impact of further flattening the z-axis metallicity gradient in the outflow. 


We observe a consistent value between the median $T_e$ in high resolution and global $T_e$ measured from the summed spectra. This implies that at a 13~pc resolution, fluctuations in $T_e$ do not affect the global measurements. We note however that our observations do not cover the entire galaxy of NGC~1569. While  \cite{Cameron2021} found a significant increase in $T_e$ at extreme ends of the major axis in Mrk~1486, for NGC~1569 we are not able to test if this is impacted by temperature fluctuations. Moreover, fluctuations in $T_e$ may be more important at small scales \citep{Peimbert2019}; we cannot test this with our data. 

We do observe local differences in $T_e$, especially in a region around SSC-B. We find that along with a much lower $T_e$, the region likewise shows significantly increased  electron density ($n_e>1000$~cm$^{-3}$). We can therefore use the differences in $T_e$ and $n_e$ to derive a local pressure of ionised gas. We find $P_{\rm HII}/k_b \sim 10^7$~cm$^{-3}$~K, which is of order 50$\times$ larger than the typical pressure of HII regions at similar metallicity \citep{McLeod2021}. We also find decreased surface brightness of H$\beta$ and H$\gamma$ and lower ionization. The region, therefore, has a lower mass surface density and is very efficient at cooling. \cite{Larsen2008} derived an age of $\sim15-25$~Myr, and \cite{Mayya2020} showed a similar finding with the absence of current Wolf-Rayet stars in the cluster. The implication is that this super star cluster is no longer experiencing a starburst, and is sufficiently old such that supernovae and early radiative feedback have already destroyed the star forming molecular clouds \citep[see the review][]{Krumholz2019}. The high pressure, high metallicity environment is consistent with a region carved out by supernova ejecta. We propose the properties of the gas around SSC-B as a possible means of identifying similar regions in other galaxies.

We have presented a comprehensive analysis of direct method observations of metal enrichment in the disk and outflow of NGC~1569. Our work is a significant step forward, as it doubles the number of observations that map the metallicity of gas in outflows. It is, however, insufficient. More observations are needed to determine the nature and validity of metal-loading and mass correlations, like we show in Fig.~\ref{fig:metal_loading}.

\section*{Acknowledgements}

Parts of this research were supported by the Australian Research Council Centre of Excellence for All Sky Astrophysics in 3 Dimensions (ASTRO 3D), through project number CE170100013.  
D.B.F. acknowledges support from Australian Research Council (ARC) Future Fellowship FT170100376 and ARC Discovery Program grant DP130101460. 
A.D.B. acknowledges partial support from AST1412419 and AST2108140. 
A.J.C. acknowledges funding from the ``FirstGalaxies'' Advanced Grant from the European Research Council (ERC) under the European Union's Horizon 2020 research and innovation programme (Grant agreement No. 789056).
R.H.-C. thanks the Max Planck Society for support under the Partner Group project "The Baryon Cycle in Galaxies" between the Max Planck for Extraterrestrial Physics and the Universidad de Concepción. R.H-C also acknowledges financial support from Millennium Nucleus NCN19058 (TITANs) and support by the ANID BASAL projects ACE210002 and FB210003.
R.R.V. and K.S. acknowledge funding support from National Science Foundation Award No. 1816462.

Some of the data presented herein were obtained at the W. M. Keck Observatory, which is operated as a scientific partnership among the California Institute of Technology, the University of California and the National Aeronautics and Space Administration. The Observatory was made possible by the generous financial support of the W. M. Keck Foundation. Observations were supported by Swinburne Keck. The authors wish to recognise and acknowledge the very significant cultural role and reverence that the summit of Maunakea has always had within the indigenous Hawaiian community. We are most fortunate to have the opportunity to conduct observations from this mountain.
\section*{Data Availability}

 All raw data files are accessible in the Keck Observatory Archive\footnote{https://www2.keck.hawaii.edu/koa/public/koa.php}.  
The data underlying this article will be shared on reasonable request to the PI, Deanne Fisher at dfisher@swin.edu.au



\bibliographystyle{mnras}
\bibliography{example} 




\appendix

\section{Emission line maps}
\label{sec:apen_emission_maps} 
\begin{figure*}
    \centering
    \includegraphics[width=1\textwidth]{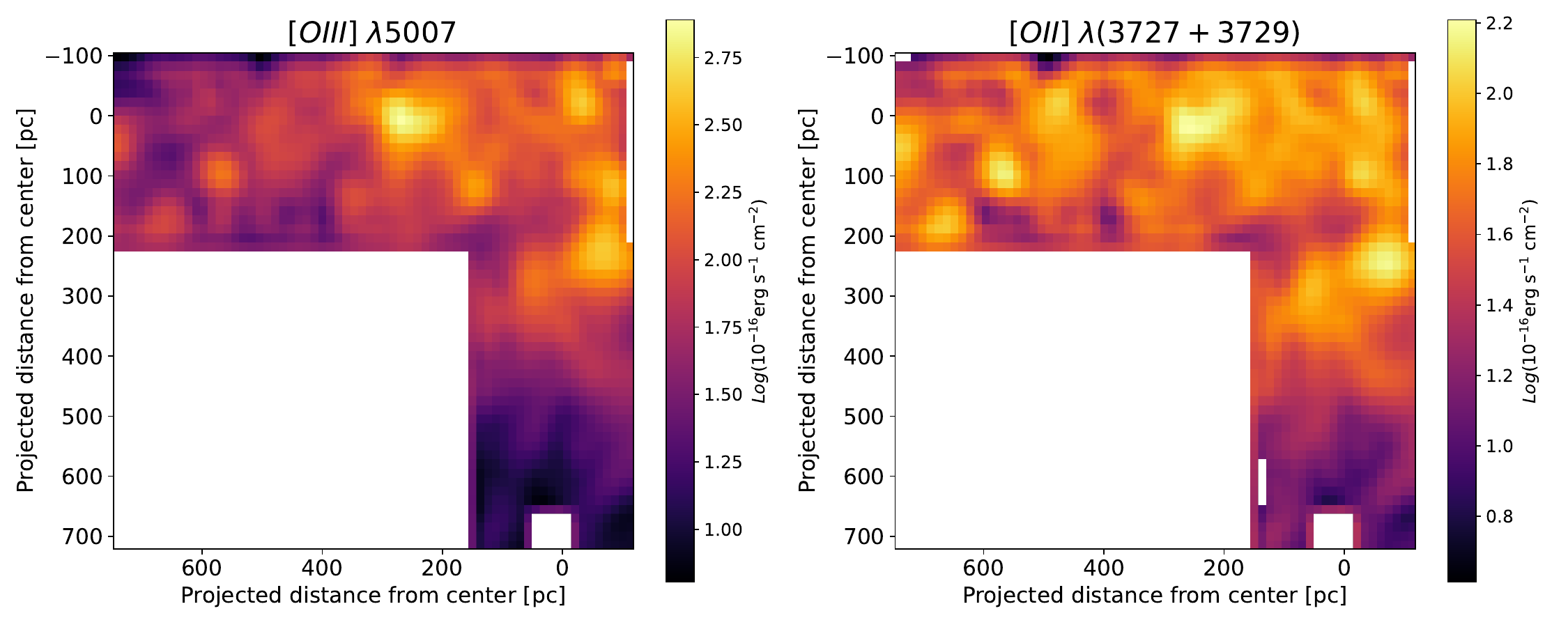}
    \caption{Left panel: Emission line flux map of [\ion{O}{III}]~$\lambda$5007 from KCWI. Right panel: Emission line flux map of [\ion{O}{II}]~$\lambda$3726 + [\ion{O}{II}]~$\lambda$3729 from KCWI. We masked out a 4$\arcsec$ $\times$ 4$\arcsec$ region around a foreground star at the position $\sim$(0, 700~pc). SSC~B is located near the emission line cavity at (0,200~pc) in the images.}
    \label{fig:emissionmapsapen}
\end{figure*}

With our Keck/KCWI observations we detect common strong emission lines at the $> 5\sigma$ level in our entire field of view. Fig.~\ref{fig:emissionmapsapen} shows the flux map for the [\ion{O}{III}]~$\lambda$5007 emission line and the total flux map of the [\ion{O}{II}]~$\lambda$3727,9 doublet. These lines are useful to determine the O32 ratio, which can be used to trace the ionization of the gas. Fig.~\ref{fig:O32_profile} shows the O32 profile along the minor axis of the galaxy, which was determined from the maps shown in Fig.~\ref{fig:emissionmapsapen}. It can be seen from both the maps and the profile that the O32 ratio decreases with increasing distance along the minor axis.

\section{Electron temperature and electron density maps}
\label{sec:apen_Te_ne_maps} 

\begin{figure*}
    \centering
    \includegraphics[width=1\textwidth]{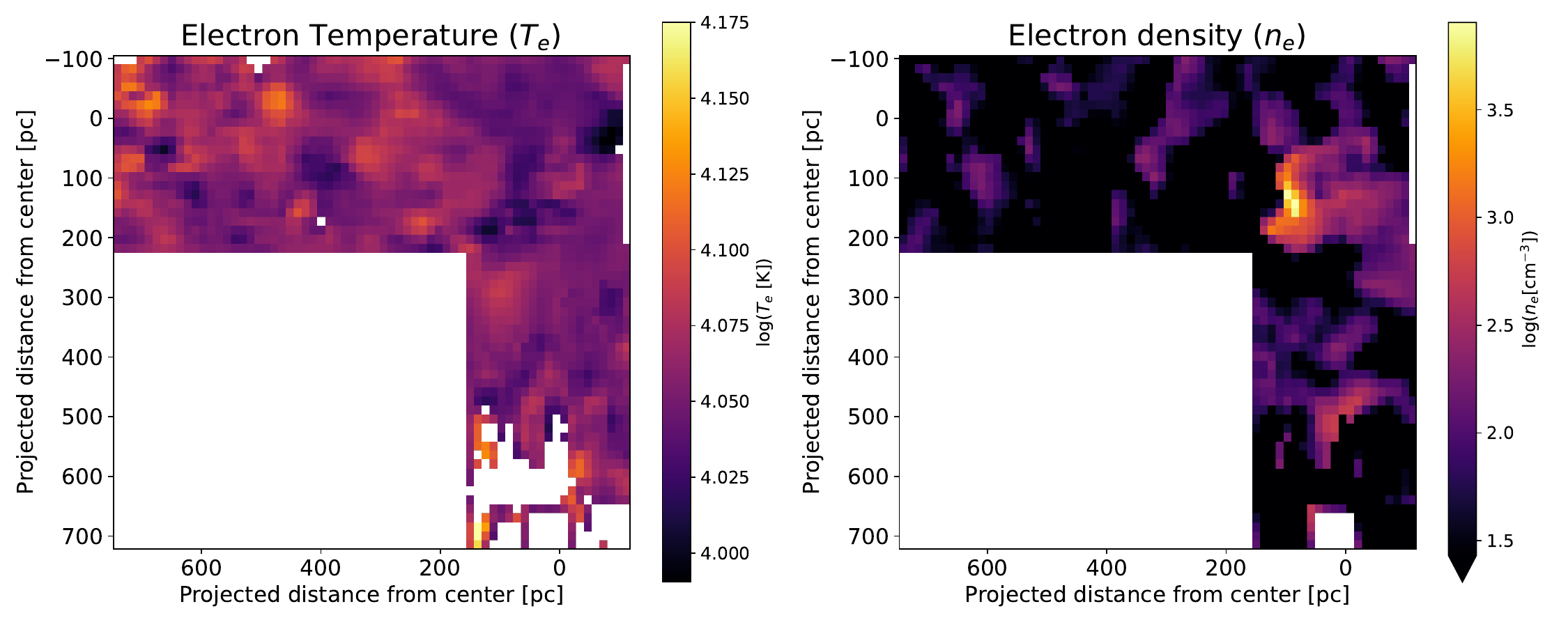}
    \caption{Left panel: Electron temperature map in Kelvins, measured from the [\ion{O}{III}]~$\lambda$4363/$\lambda$5007 ratio. Right panel: Electron density map, measured from the [\ion{O}{II}]~$\lambda$3729/$\lambda$3726 ratio. Black spaxels show positions at which the $\lambda$3729/$\lambda$3726 ratio reaches the low density limit.}
    \label{fig:Te_ne_map}
\end{figure*}

We measure the electron temperature of the gas from the [\ion{O}{III}]~$\lambda$4363/$\lambda$5007 ratio. Fig.~\ref{fig:Te_ne_map} shows the $T_e$ map. $T_e$ is 
inversely proportional to the metallicity, shown in Fig.~\ref{fig:Te_Z_maps}. From the [\ion{O}{II}]~$\lambda$3729/$\lambda$3726 we can measure the electron density. Fig.~\ref{fig:Te_ne_map} show our measured electron density map. This ration can be used to measure $n_e$ for values of $0.347 < [\ion{O}{II}] \lambda 3729 / \lambda 3726  < 1.5$. In Fig.~\ref{fig:Te_ne_map} we plotted in grey all spaxels for which the ratio, within its errors, is outside of the range. The lowest value of $n_e$ that we measure is $\sim$26~cm$^{-3}$.

\bsp	
\label{lastpage}
\end{document}